\documentclass[usenatbib,letterpaper]{mn2e}

\usepackage{graphics}
\usepackage{epsfig}
\usepackage{natbib}

\newcommand{\kms}{km~s\ensuremath{^{-1}}}
\newcommand{\msun}{$M_{\odot}$}
\newcommand{\nuvr}{NUV$-r$}
\newcommand{\mh}{$M_{H_2}$}

\newcommand{\mstar}{$M_{\ast}$}
\newcommand{\must}{$\mu_{\ast}$}
\newcommand{\hmol}{$H_2$}
\newcommand{\rmol}{$R_{mol}$}
\newcommand{\nodata}{$...$}

\newcommand{\ntot}{222}
\newcommand{\ndet}{119}
\newcommand{\nondet}{103}
\newcommand{\noff}{25}
\newcommand{\nred}{68}

\newcommand{\apj}{ApJ}
\newcommand{\apjs}{ApJS}
\newcommand{\apjl}{ApJ}
\newcommand{\aj}{AJ}
\newcommand{\mnras}{MNRAS}
\newcommand{\aap}{A\&A}
\newcommand{\aaps}{A\&AS}
\newcommand{\araa}{ARA\&A}
\newcommand{\pasj}{PASJ}
\newcommand{\pasp}{PASP}
\newcommand{\nat}{Nature}

\begin{document}

\title[COLD GASS: Molecular Gas in Massive Galaxies]{COLD GASS, an IRAM Legacy Survey of Molecular Gas in Massive Galaxies: 
I. Relations between H$_2$, HI, Stellar Content and Structural Properties}

\author[A. Saintonge et al.]{Am\'{e}lie Saintonge$^{1,2}$\thanks{E-mail: amelie@mpe.mpg.de}, Guinevere Kauffmann$^{1}$,  Carsten Kramer$^{3}$,  Linda J. Tacconi$^{2}$,  
\newauthor
Christof Buchbender$^{3}$, Barbara Catinella$^{1}$,  Silvia Fabello$^{1}$, Javier Graci\'{a}-Carpio$^{2}$, 
\newauthor
Jing Wang$^{1,4}$, Luca Cortese$^{5}$, Jian Fu$^{6,1}$, Reinhard Genzel$^{2}$, Riccardo Giovanelli$^{7}$, 
\newauthor
Qi Guo$^{8,9}$, Martha P. Haynes$^{7}$, Timothy M. Heckman$^{10}$, Mark R. Krumholz$^{11}$, 
\newauthor
Jenna Lemonias$^{12}$, Cheng Li$^{6,13}$,  Sean Moran$^{10}$, Nemesio Rodriguez-Fernandez$^{14}$, 
\newauthor
David Schiminovich$^{12}$, Karl Schuster$^{14}$ and Albrecht Sievers$^{3}$ \\
$^{1}$Max-Planck Institut f\"{u}r Astrophysik, 85741 Garching, Germany\\
$^{2}$Max-Planck Institut f\"{u}r extraterrestrische Physik, 85741 Garching, Germany\\
$^{3}$Instituto Radioastronom\'{i}a Milim\'{e}trica, Av. Divina Pastora 7, Nucleo Central, 18012 Granada, Spain\\
$^{4}$Center for Astrophysics, University of Science and Technology of China, 230026 Hefei, China\\
$^{5}$European Southern Observatory, Karl-Schwarzschild-Str. 2, 85748 Garching, Germany\\
$^{6}$Key Laboratory for Research in Galaxies and Cosmology, Shanghai Astronomical Observatory, Chinese Academy of Sciences, \\ 
Nandan Road 80, Shanghai 200030, China\\
$^{7}$Center for Radiophysics and Space Research, Cornell University, Ithaca, NY 14853, USA\\
$^{8}$National Astronomical Observatories, Chinese Academy of Sciences, Beijing 100012, China\\
$^{9}$Institute for Computational Cosmology, Department of Physics, Durham University, South Road, Durham DH1 3LE, UK\\
$^{10}$Johns Hopkins University, Baltimore, Maryland 21218, USA\\
$^{11}$Department of Astronomy and Astrophysics, University of California, Santa Cruz, CA 95064, USA\\
$^{12}$Department of Astronomy, Columbia University, New York, NY 10027, USA\\
$^{13}$Max-Planck-Institut Partner Group, Shanghai Astronomical Observatory\\
$^{14}$Institut de Radioastronomie Millim\'{e}trique, 300 Rue de la piscine, 38406 St Martin d'H\`{e}res, France}

\maketitle

\begin{abstract}
We are conducting COLD GASS, a legacy survey for molecular gas in nearby galaxies.  Using the IRAM 30m telescope, we measure the CO(1-0) line in a sample of  $\sim$ 350 nearby ($D_L\simeq100-200$ Mpc), massive galaxies ($\log(M_{\ast}/M_{\odot})>10.0$).  The sample is selected purely according to stellar mass, and therefore provides an unbiased view of molecular gas in these systems. By combining the IRAM data with SDSS photometry and spectroscopy, GALEX imaging and high-quality Arecibo HI data, we investigate the partition of condensed baryons between stars, atomic gas and molecular gas in $0.1-10L^{\ast}$ galaxies.  In this paper, we present CO luminosities and molecular hydrogen masses for the first \ntot\ galaxies.  The overall CO detection rate is 54\%, but our survey also uncovers the existence of sharp thresholds in galaxy structural parameters such as stellar mass surface density and concentration index,  below which all galaxies have a measurable cold gas component but above which the detection rate of the CO line drops suddenly. The mean molecular gas fraction $M_{H2}/M_{\ast}$ of the CO detections is $0.066\pm0.039$, and this fraction does not depend on stellar mass, but is a strong function of \nuvr\ colour.  Through stacking, we set a firm upper limit of $M_{H2}/M_{\ast}=0.0016\pm0.0005$ for red galaxies with \nuvr$>5.0$. The average molecular-to-atomic hydrogen ratio in present-day galaxies is 0.3, with significant scatter from one galaxy to the next.    The existence of strong detection thresholds in both the HI and CO lines suggests that ``quenching'' processes have occurred in these systems. Intriguingly, atomic gas strongly dominates in the minority of galaxies with significant cold gas that lie above these thresholds. This suggests that some re-accretion of gas may still be possible following the quenching event.   
\end{abstract}


\section{Introduction}
\label{intro}

Perhaps the most fascinating aspect of nearby galaxies is the intricately interwoven system of correlations between their global properties. These correlations form the basis of the so-called ``scaling laws",  which are fundamental because they provide a quantitative means of characterizing how  the physical properties of galaxies relate to each other. Galaxy scaling relations also provide the route to understanding the internal physics of galaxies, as well as their formation and evolutionary histories. 

We currently enjoy a diverse array of scaling laws that describe the stellar components of galaxies, for example the Tully-Fisher relation for spiral galaxies \citep{tf77}, and the fundamental plane for ellipticals \citep{jorgensen96}. Both relations provide important constraints on how these systems have assembled.  However, few well-established scaling laws exist describing how the cold gas is correlated with the other global physical properties of galaxies. The only well-studied relation is the Schmidt-Kennicutt star formation law \citep{kennicutt98a}, relating the formation rate of new stars and the surface density of cold gas in disks. 

The reason why so few scaling laws involving cold gas and global galaxy properties, such as masses, sizes and bulge-to-disk ratios, exist in the literature, is the difficulty in acquiring suitable data. There are four general requirements on the data if the derived scaling laws are to be reliable: (1) homogeneous and accurate measurements of all the physical properties under consideration, (2) unbiased measurement of every property with respect to every other property, (3)  sample selection that ensures large dynamic range of the various physical properties under consideration, (4) a large enough sample to define both the mean relation and the scatter about the mean. As we will describe, existing data sets do not, in general, meet all of these conditions.  

Line emission from the CO molecule was first detected in the central parts and disks of nearby galaxies 35 years ago \citep{rickard75,solomon75,combes77}.  A decade later, CO measurements existed for $\sim100$ galaxies \citep[see the compilation of][]{verter85}, and in the following years, several larger systematic studies of molecular gas in nearby galaxies were performed. The largest effort was the FCRAO Extragalactic CO Survey \citep{young95}, which measured the CO $J=1\rightarrow0$ line (hereafter, CO(1-0)) in 300 nearby galaxies.  Because of a strong correlation between infrared luminosity and CO luminosity \citep[e.g.][]{sanders91,sanders96}, a significant part of this early work was done targeting luminous infrared galaxies \citep[e.g.][]{radford91,solomon97}.  Even the FCRAO Survey, still considered as the reference for CO measurements in the nearby universe, targeted galaxies selected on infrared  or $B-$band luminosity.  

Recognizing this bias toward ``exceptional" galaxies (e.g. starbursts and interacting systems), \citet{braine93} observed the CO $1\rightarrow0$ and $2\rightarrow1$ rotational transition lines with the IRAM 30m telescope for a magnitude-limited sample of 81 normal spiral galaxies.   Other attempts at measuring molecular gas in normal galaxies include the work of \citet{kenney88} for Virgo cluster spirals, of \citet{sage93} for nearby non-interacting spirals, and of \citet{boselli97} for Coma cluster spirals.  

These pioneering studies constrained molecular gas properties in nearby galaxies as a function of morphology \citep[e.g.][]{wiklind89,thronson89}, star formation rate or infrared luminosity \citep[e.g.][]{sanders85,gao04}, atomic gas contents \citep[e.g.][]{young89a}, environment \citep[e.g.][]{kenney89,casoli91,boselli97}, and for resolved studies, position within galaxy disks.  Highlights from these studies include the observations that molecular gas distributions decline monotonically with galaxy-centric radius unlike the atomic gas distributions, that IR-luminous galaxies are also CO-bright, with molecular gas concentrated within the inner kpc of these mostly interacting systems, and that the total gas mass fraction as well as the molecular-to-atomic ratio are functions of Hubble type. 

Nevertheless, most of the samples did not meet all of the criteria listed above that would allow for accurate scaling laws to be derived; some samples were biased towards a particular galaxy type (e.g. infrared-bright objects), some of the more unbiased samples did not cover enough parameter space (e.g. targeting only spiral galaxies), some samples suffered from aperture problems, some samples were too small, and attempts to combine different  samples to remedy these problems led to inhomogeneous datasets (see also \S \ref{archive}).   

Recently, much effort has been put into obtaining homogeneous and relatively deep high spatial resolution molecular gas maps covering the optical disks of nearby galaxies \citep{regan01,kuno07,leroy09}. These samples are excellent for studying star formation laws within galaxies \citep[e.g.][]{bigiel08}, but the number of objects is too small to adequately define global scaling relations. 

With reliable measurements of molecular gas for a large, unbiased sample of galaxies, it is possible not only to quantify scaling relations, but also to construct an accurate molecular gas mass function.  Current estimates are based on inherently inhomogeneous samples \citep{keres03,obreschkow09}.  We can also investigate the molecular gas properties of galaxy samples for which dedicated surveys do exist, but where the number of objects studied has been very small, for example early-type galaxies \citep[e.g.][]{combes07,krips10}, and galaxies with active nuclei \citep[e.g.][]{helfer93,sakamoto99,garcia03}.  A large unbiased sample of galaxies which can serve as a reference for such particular objects would also be very valuable.

In this paper, we introduce COLD GASS, a new survey for molecular gas in nearby galaxies.  Upon completion, it will have measured fluxes in the CO(1-0) line for a purely mass-selected sample of at least 350 galaxies.  The sample  contains galaxies with a wide range of Hubble types from star-forming spirals to ``red and dead" ellipticals.  With its new, large-bandwidth receivers, the IRAM 30m telescope is the instrument of choice to conduct a new large molecular gas survey, allowing the community to move from dedicated studies of particular types of galaxies,  to larger systematic efforts.  COLD GASS will provide a definitive, unbiased census of the partition of condensed baryons in the local Universe into stars, atomic and molecular gas in galaxies covering over two orders of magnitude in luminosity.  

In \S \ref{sample}-\ref{iramdata}, we present an overview of the survey and of the sample selection, and describe the CO measurements and ancillary datasets. In \S \ref{scalrel} and \ref{compHI} we present the first COLD GASS scaling relations, correlating molecular gas masses with  global galaxy parameters including stellar mass and atomic gas mass.  Throughout the paper, distance-dependent quantities are calculated for a standard flat $\Lambda$CDM cosmology with $H_0=70$\kms\ Mpc$^{-1}$, and we adopt a conversion factor from CO luminosity to $H_2$ mass of $\alpha_{CO}=3.2$\msun\ (K \kms\ pc$^2$)$^{-1}$ (which does not account for the presence of Helium), unless otherwise specified.

\section{Survey description and Sample selection}
\label{sample}

The conditions required to obtain reliable scaling laws and listed in \S \ref{intro} are routinely met by optically-selected samples of galaxies at low redshift. The Sloan Digital Sky Survey \citep[SDSS;][]{sdss} with its 5-band optical imaging campaign over a quarter of the sky and its follow-up spectroscopy of close to a million galaxies has facilitated the study of galaxy scaling relations studies at an unprecedented level of detail.

At radio wavelengths,  a series of large blind HI surveys have become possible thanks to a number of new multi-feed arrays.  The most advanced of these, the Arecibo Legacy Fast ALFA Survey \citep[ALFALFA;][]{ALFALFA1}, will have detected upon completion $\sim 30,000$ galaxies out to distances of $\sim 200$Mpc. Although ALFALFA measurements are accurate, homogeneous, and unbiased, the survey is shallow, with the result that it does not probe a large dynamic range in HI-to-stellar mass ratio for all but the very nearest galaxies. For example, in the redshift range $0.025<z<0.05$, the median value of $M_{HI}/M_{\ast}$ for ALFALFA detections with $M_{\ast}>10^{10}$\msun\ is $\sim25\%$. 

\subsection{GASS}
\label{gass}

To overcome this issue, the {\it GALEX} Arecibo SDSS Survey \citep[GASS;][]{GASS1} was designed to measure the neutral hydrogen content for a large, unbiased sample of $\sim1000$ massive galaxies ($M_{\ast}>10^{10}$ \msun), via longer pointed observations. GASS is a large program currently under way at the Arecibo 305m telescope, and is producing some of the first unbiased atomic gas scaling relations in the nearby universe \citep{GASS1,GASS2,fabello10}. 

Details about the GASS survey design, target selection, and observing procedures are given in \citet{GASS1}.  In short, the galaxies observed as part of GASS are selected at random out of a larger parent sample of galaxies that meet the following criteria:
\begin{enumerate}
\item They are located within the area of overlap of the SDSS spectroscopic survey, the ALFALFA survey, and the projected footprint of the {\it GALEX} Medium Imaging Survey (MIS).
\item They lie in the redshift range $0.025<z<0.05$. 
\item They have a stellar mass in the range $10^{10}<M_{\ast}/M_{\odot}<10^{11.5}$.
\end{enumerate}
The GASS sample is selected out of this parent sample to produce a flat $\log M_{\ast}$ distribution. No other selection criteria on colour, morphology, or spectral properties for example are applied.   This sample therefore provides us with a complete picture of how the cold atomic gas relates to other properties such as stellar mass, luminosity, stellar surface mass density and colour.  

\citet{GASS1} present the first GASS data release, which includes $\sim20\%$ of the final sample. They show that there exist strong anti-correlations between the atomic gas mass fraction and stellar mass, stellar mass surface density and \nuvr\ colour.  GASS also aims at studying the galaxies that are transitioning between a blue, star-forming state and a red passive state (and vice-versa). These are identified as outliers from the mean scaling relations.  The ultimate goal is to understand the physical processes that affect the gas content of these galaxies (e.g. accretion or quenching) and in turn the star formation process.

\subsection{COLD GASS}
\label{sampleselect}

\begin{figure}
\includegraphics[width=84mm]{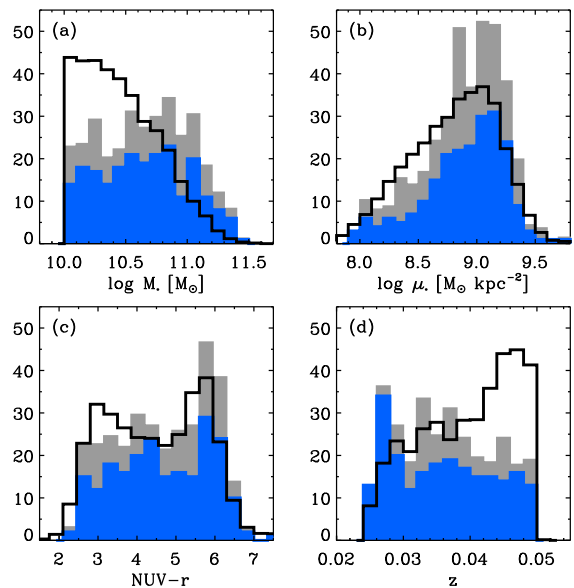}
\caption{Distribution of sources observed as of 25 Oct. 2010 (filled blue histograms), compared to the proposed final COLD GASS sample (filled gray histograms).  The solid line in each panel shows the distribution of objects in the superset of $\sim$12000 galaxies out of which the GASS sample is extracted, scaled down to the number of objects in the COLD GASS master list (350).   The GASS sample is selected as to produce a flat distribution in $\log M_{\ast}$. \label{distribs}}
\end{figure}

We are in the process of constructing a CO Legacy Database for the GASS survey (COLD GASS), measuring the molecular gas content of a significant subsample of the GASS galaxies.  We will then be able to quantify the link between atomic gas,  molecular gas and stars in these systems. 

COLD GASS is designed to meet all the requirements to produce reliable gas scaling relations:
\begin{enumerate}
\item Galaxies in our redshift range ($0.025<z<0.05$) have angular diameters that are small enough to enable accurate measurement of the total CO line flux with a single IRAM 30m pointing for the majority (80\%) of the galaxies.  For the remaining objects, we recover the total flux by adding a single offset pointing (see \S \ref{offsets} for details). 
\item In the mass range we study ($10^{10}<M_{\ast}/M_{\odot}<10^{11.5}$), the metallicity of the galaxies is around solar \citep{tremonti04}. The CO line flux therefore provides a reasonably accurate measurement of the total molecular gas content using a single conversion factor, $\alpha_{CO}$.
\item The $\sim 350$ targets are selected at random from the GASS survey, the sample is therefore unbiased. 
\item We integrate until the CO line is detected, or until we reach an upper limit in molecular gas mass to stellar mass ratio ($f_{H_2}\equiv M_{H_2}/M_{\ast}$) of $\sim1.5\%$.
\item Upon completion of the survey, the sample size of  at least 350 galaxies will be large enough to determine accurately a set of scaling laws involving three parameters and to measure the scatter around these relations.
\end{enumerate}

Distributions of some basic parameters of the COLD GASS sample are  shown in Figure \ref{distribs} and compared to a purely volume-limited superset of galaxies.  The COLD GASS targets are selected randomly from the GASS survey and therefore share the flat $\log M_{\ast}$ distribution designed for that survey to ensure even sampling of the stellar mass parameter space.  As seen in  Figure \ref{distribs}, a purely volume-limited sample is richer in low mass galaxies, and poorer in high mass systems. The uniform mass distribution also has the effect of flattening the colour distribution and reducing the number of galaxies with low stellar mass surface densities.  We note that when deriving scaling relations and calculating sample averages, we statistically correct for this ``mass bias",  as done by \citet{GASS1} for the GASS sample (see \S \ref{scalrel}) .

\section{SDSS, {\it GALEX} and Arecibo Observations}
\label{data}

\subsection{Optical and UV measurements}

Parameters such as redshifts, sizes, magnitudes, and Galactic extinction factors are retrieved from the database of SDSS DR7 \citep{DR7}.  The UV data are taken from the {\it GALEX} All-sky and Medium Imaging surveys \citep[AIS and MIS, respectively, see][]{martin05}. 

The SDSS and {\it GALEX} images are reprocessed following \citet{wang10}, in order to obtain accurate aperture photometry.  The process includes registering the images and smoothing them to a common PSF.  The SDSS $r-$band images are convolved to the resolution of the UV imaging before Sextractor is used to calculate magnitudes in consistent apertures,  therefore ensuring that measurements in different bands represent similar physical regions of the galaxies.  The derived \nuvr\ colours are corrected for Galactic extinction using the prescription of \citet{wyder07} \citep[see also][]{GASS1}.

Stellar masses are calculated from the SDSS photometry using the SED-fitting technique of \citet{salim07}, assuming a Chabrier IMF \citep{chabrier03}.   A variety of model SEDs from the \citet{bc03} library are fitted to each galaxy, building a probability distribution for its stellar mass.  The stellar mass assigned to a galaxy is then the mean of this distribution, while the measurement error is estimated from its width.  The systematic uncertainty between different technique to derive photometric stellar masses from SDSS measurements is $<0.1$dex, as estimated by \citet{dutton11}.

The main optical- and UV-derived parameters used throughout this paper are presented in Table \ref{params}, for all galaxies within the present COLD GASS data release.   Column 1 and 2 give the GASS and SDSS ID numbers, respectively, column 3 gives the optical redshift from SDSS spectroscopy, while column 4 lists the stellar mass and column 5 the stellar mass surface density, which we calculate as:
\begin{equation}
\mu_{\ast}=\frac{M_{\ast}}{2\pi R_{50,z}^2},
\end{equation}
where $R_{50,z}$ is the $z-$band 50\% flux intensity petrosian radius, in kiloparsecs.   In column 6, we give the $g-$band optical diameter ($D_{25}$), and in column 7 the concentration index ($C\equiv R_{90}/R_{50}$, where $R_{50 }$ and $R_{90}$ are from $r-$band photometry). Finally, columns 8 and 9 present the \nuvr\ colour and the $r-$band model magnitude, both corrected for Galactic extinction. 

\begin{table*}
\begin{minipage}{150mm}
\caption{Optical and UV parameters of the COLD GASS galaxies$^a$}
\label{params}
\begin{tabular}{ccccccccc}
\hline
GASS ID & SDSS ID & $z_{SDSS}$ & 
$M_{\ast}$ & $\mu_{\ast}$ &
$D_{25}$ & $R_{90}/R_{50}$ & NUV$-r$ & $r$ \\
 &  &  & $[\log M_{\odot}]$ & $[\log M_{\odot} \rm{kpc}^{-2}]$ & [$\arcsec$] & 
 & [mag] & [mag] \\
\hline
 11956 &    J000820.76+150921.6 & 0.0395 & 10.09 & 8.48 &     22.5 & 2.15 &     3.04 & 16.28 \\
 12025 &    J001934.54+161215.0 & 0.0366 & 10.84 & 9.13 &     34.3 & 3.03 &     5.93 & 14.73 \\
 12002 &    J002504.00+145815.2 & 0.0367 & 10.48 & 9.41 &     24.2 & 3.17 &     6.25 & 15.46 \\
 11989 &    J002558.89+135545.8 & 0.0419 & 10.69 & 9.18 &     23.7 & 3.02 &     5.79 & 15.13 \\
 27167 &    J003921.66+142811.5 & 0.0380 & 10.37 & 9.14 &     21.1 & 2.77 &     4.48 & 15.49 \\
  3189 &    J004023.48+143649.4 & 0.0384 & 10.05 & 7.92 &     37.7 & 1.96 &     2.77 & 15.65 \\
  3261 &    J005532.61+154632.9 & 0.0375 & 10.08 & 8.57 &     22.8 & 2.54 &     2.63 & 15.48 \\
  3318 &    J010238.29+151006.6 & 0.0397 & 10.53 & 8.98 &     26.4 & 3.05 &     5.73 & 15.21 \\
  3439 &    J010905.96+144520.8 & 0.0386 & 10.35 & 8.78 &     32.5 & 2.90 &     3.05 & 15.48 \\
  3465 &    J011221.82+150039.0 & 0.0292 & 10.19 & 8.93 &     28.7 & 2.89 &     3.63 & 15.33 \\
  3645 &    J011501.75+152448.6 & 0.0307 & 10.33 & 8.93 &     28.1 & 2.71 &     3.97 & 15.11 \\
  3509 &    J011711.65+132027.3 & 0.0484 & 10.81 & 9.18 &     31.1 & 3.11 &     4.14 & 15.27 \\
  3519 &    J011728.11+144215.9 & 0.0427 & 10.74 & 8.64 &     34.2 & 2.20 &     3.68 & 14.94 \\
  3505 &    J011746.76+131924.5 & 0.0479 & 10.21 & 8.83 &     17.7 & 3.30 &     4.92 & 16.35 \\
  3504 &    J011823.44+133728.4 & 0.0380 & 10.16 & 7.91 &     37.7 & 1.84 &     2.85 & 15.34 \\
\hline
\end{tabular}
$^a$ Table 1 is published in its entirety in the electronic edition of the journal.  A portion is shown here as example of its format and contents.
\end{minipage}
\end{table*}

\subsection{HI masses}

Details of the HI observations are described in \citet{GASS1}, so we only provide a brief overview here. The survey builds upon existing HI databases: the Cornell digital HI archive \citep{springob05} and the ALFALFA survey \citep{ALFALFA1}.   HI data for about 20\% of the GASS sample (the most gas-rich objects), can be found in either of these sources.  For the rest of the sample, observations are carried out at the Arecibo Observatory.  Integration times are set such as to detect HI gas mass fractions ($f_{HI}=M_{HI}/M_{\ast}$) of 1.5\% or more.  Observations are carried out using the $L-$band Wide receiver and the interim correlator, providing coverage of the full frequency interval of the GASS targets at a velocity resolution of 1.4 \kms \ before smoothing.  Data reduction includes Hanning smoothing, bandpass subtraction, radio frequency interference (RFI) excision, flux calibration and weighted combination of individual spectra.  Total HI-line fluxes, velocity widths and recessional velocities are then measured using linear fitting of the edges of the HI profiles \citep[e.g.][]{springob05,catinella07}.

\section{IRAM Observations}
\label{iramdata}

\subsection{Observing procedure \label{obs_setup}}

Observations are carried out at the IRAM 30m telescope.  We use the Eight Mixer Receiver (EMIR) to observe the CO(1-0) line (rest frequency, 115.271 GHz) .  The CO $1\rightarrow0$ transition traces well the entire molecular gas contents of the galaxies at $n(H_2)>10^2$ cm$^{-3}$ \citep[see e.g. the Appendix in][]{tacconi08}.  In the 3mm band (E090), EMIR offers two sidebands with 8 GHz instantaneous bandwidth per sideband and per polarisation. With a single tuning of the receiver at a frequency of 111.4081 GHz, we are able to detect the redshifted CO(1-0) line for all the galaxies in our sample ($0.025<z<0.05$), within the 4 GHz bandwidth covered by the correlators.  This single tuning procedure  results in enormous time savings of 15 minutes per source, and in an improved relative calibration accuracy.  Also, the frequency range covered benefits from a considerably improved atmospheric transmission as compared to the CO(1-0) rest frequency.  The second band is tuned to a frequency of 222.8118 GHz (E230 band), to cover the redshifted CO(2-1) line which falls within the available 4 GHz bandwidth for about 75\% of our sample. We postpone the presentation and analysis of the CO(2-1) data until the survey is completed, to maximize the sample size. 

The wobbler-switching mode is used for all the observations with a frequency of 1Hz and a throw of 180 \arcsec.  The Wideband Line Multiple Autocorrelator (WILMA) is used as the backend, covering 4 GHz in each linear polarisation, for each band.  WILMA gives a resolution of 2 MHz ($\sim 5$ \kms\ for the 3mm band).  We also simultaneously record the data with the 4MHz Filterbank, as a backup. 

Observations for this first data release were conducted between 2009 December and 2010 October.  Atmospheric conditions varied greatly, with an average of 6mm of precipitable water vapor (PWV).   We also fold into this catalog 15 galaxies observed in 2009 June, as part of a pilot program designed to test the feasibility of the survey.  These galaxies were selected to be HI-rich ($M_{HI}/M_{\ast}>0.1$), but this selection bias does not affect the overall sample.

\subsection{Observing strategy}
\label{obsstrat}

\begin{figure}
\includegraphics[width=84mm]{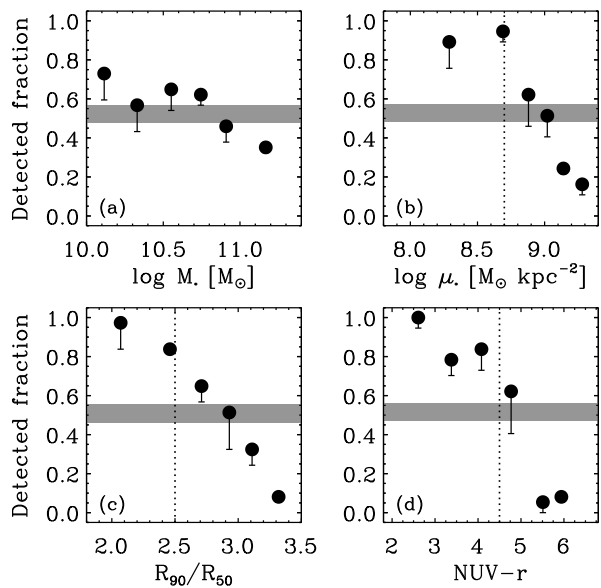}
\caption{Fraction of galaxies with a detection in the CO(1-0) line as a function of (a) stellar mass, (b) stellar mass surface density, (c) concentration index (defined as $R_{90}/R_{50}$, the ratio between the $r-$band radii encompassing 90\% and 50\% of the light), and (d) \nuvr\ colour.   The results are shown in equally populated bins, each containing 37 galaxies. The gray shaded region shows the overall detection rate of $53.6\%$, down to $47.3\%$ if all tentative detections are excluded.   The downward error bars show the effect of excluding tentative detections in each individual bin. In panels (b-d), the vertical dotted line indicates the critical value where the detection rate suddenly drops below $\sim80\%$. \label{detfrac}}
\end{figure}

Observations are carried out in fixed observing blocks and as poor-weather backups for higher frequency programs.  We accommodate to the changing weather conditions by making ``real time" decisions on targets.  We observe the bluer galaxies (generally CO-luminous) under poorer weather conditions.  These galaxies require on average an rms sensitivity of 1.7 mK per 20 \kms-wide channel to achieve a reliable detection of the CO(1-0) line with $S/N>5$.  As seen in Figure \ref{detfrac}d, galaxies with colour \nuvr$<4.5$ have a detection rate greater than 80\%.  When the atmospheric water vapor level is low, we preferentially observe the redder galaxies, which have a very low detection rate (Fig. \ref{detfrac}d).  In order to set firm upper limits for these galaxies, we require low noise values to reach our integration limit of $M_{H2}/M_{\ast}=1.5\%$.  Under good observing conditions, sensitivity to this minimum gas fraction, or an absolute minimum rms of 1.1mK (per 20\kms-wide channel), is reached within 1-1.5 hour.  This absolute minimum rms is imposed in order to keep the integration time per galaxy $<2$ hours, and translates in a detection limit that is higher than the nominal value of 1.5\% for galaxies with $\log M_{\star}/M_{\odot}<10.6$.

\begin{figure*}
\begin{minipage}{165mm}
\includegraphics[width=165mm]{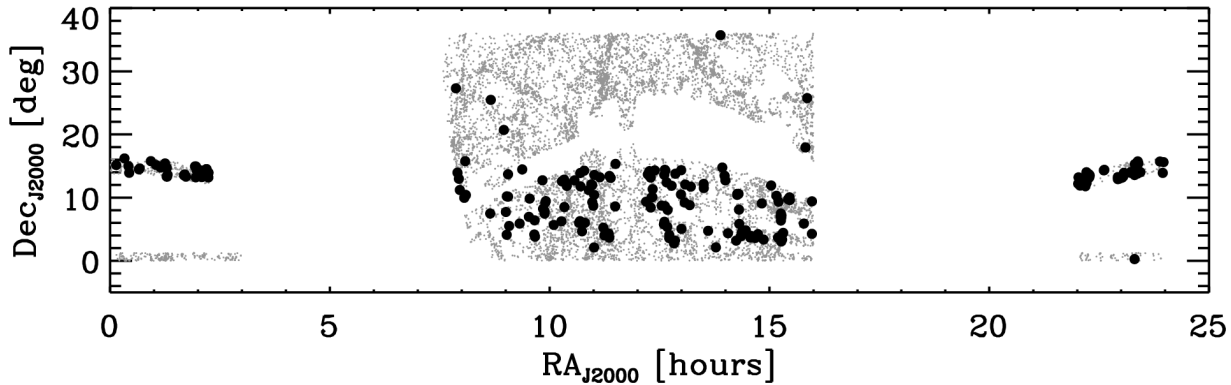}
\caption{Sky distribution of the COLD GASS sample.  Galaxies are selected in the area of intersection of the SDSS, GALEX MIS and ALFALFA HI surveys.  Gray dots show all 12000 galaxies satisfying the GASS selection criteria (see \S \ref{gass}) within the SDSS+GALEX footprints, and the circles represent the \ntot\ COLD GASS galaxies that are part of this data release.  \label{sky}}
\end{minipage}
\end{figure*}

The efficiency of the observations is also maximized by our single tuning approach (see \S \ref{obs_setup}), and by the fact that the galaxies are mostly concentrated in a declination strip with $0^{\circ} < \delta < +15^{\circ}$, as shown in Figure \ref{sky}.   Not only does this allow us to move quickly from one source to the next without repeating pointing correction measurements, it makes it possible to almost always observe at elevations larger than 45$^{\circ}$, minimizing atmospheric opacity. 

\subsection{Data reduction}
\label{linemes}

The data are reduced with the CLASS software.  All scans are visually examined, and those with distorted baselines, increased noise due to poor atmospheric conditions, or anomalous features are discarded.  The individual scans for a single galaxy are baseline-subtracted (first order fit) and then combined. This averaged spectrum is finally binned to a resolution of $\sim 20$\kms,  and the standard deviation of the noise per such channel is recorded ($\sigma_{rms}$). 

Flux in the CO(1-0) line is measured by adding the signal within an appropriately defined windowing function.  If the line is detected, the window is set by hand to match the observed line profile.  If the CO line is undetected or very weak, the window is set either to the full width of the HI line ($W50_{HI}$) or to a width of 300 \kms\ in case of an HI non-detection.  For the non-detection, an upper limit for the flux of $5\epsilon_{obs}$ (see Eq. \ref{errmes}) is set.

The central velocity and total width of the detected CO lines are then measured using a custom-made IDL interactive script.  The peaks of the signal are identified, and a linear fit is applied to each side of the profile between the 20\% and 80\% peak flux level.  The width of the line, $W50_{CO}$, is then measured as the distance between the points on each of the fits corresponding to 50\% of the peak intensity.  The recession velocity is taken as the midpoint of this line. This method is described in \citet{springob05} and \citet{catinella07}, and is also used to measure the HI line-widths of the GASS sample.

\subsection{Aperture Corrections }
\label{offsets}

\begin{figure}
\includegraphics[width=84mm]{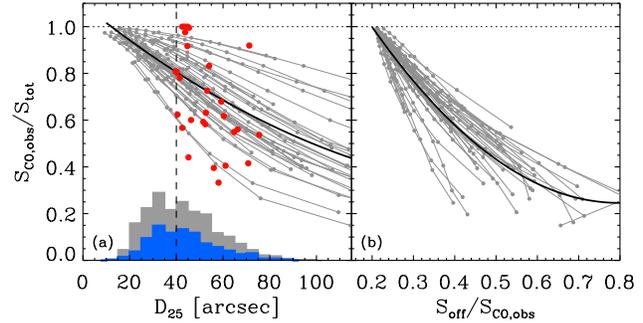}
\caption{Aperture corrections for the IRAM 30m observations of the COLD GASS sample. (a): Ratio between the flux recovered by a central pointing with the IRAM 22\arcsec\ beam to the total flux of the galaxy ($S_{CO,obs}/S_{tot}$), as a function of optical diameter ($D_{25}$).  Small gray points indicate the results from the simulated observations of a sample of 40 nearby spiral galaxies with high quality CO(1-0) maps \citep{kuno07} when they are placed at various redshifts in the range 0.01-0.05 (each set of points connected by a broken line corresponds to one of the galaxies, placed at different redshifts).  A single central pointing recovers at least 60\% of the flux in galaxies with $D_{25}<40\arcsec$.   The histogram indicates the size distribution of the COLD GASS sample (with the distribution of blue galaxies, \nuvr$<$4.3, indicated in blue). (b): With an additional offset observation 0.75 beam (i.e. $16\arcsec$) from the centre of the galaxy along the major axis ($S_{off}$), it is possible to derive the fraction of CO(1-0) line flux detected at the center and from that estimate the total molecular gas mass in galaxies with $D_{25}>40\arcsec$. The large red symbols in panel (a) indicate the aperture corrections estimated using this method for the \noff\ COLD GASS galaxies for which we performed offset pointings to date, and the dashed line is the size threshold (40\arcsec) for a galaxy to require an offset pointing. \label{figapcor}}
\end{figure}

Because the galaxies targeted are at a distance of at least 100 Mpc ($z>0.025$), most of them can be observed with a single pointing of the IRAM 30m, which has a beam with a FWHM of 22\arcsec\ at a wavelength of 3mm.  However some of the galaxies have optical diameters in excess of this, and an aperture correction needs to be applied. 

We derive aperture corrections using a set of nearby galaxies with accurate CO maps \citep{kuno07}. We simulate the impact of observing galaxies with the IRAM beam by taking each of the maps, placing it at different redshifts in the range $0.025<z<0.05$, and computing the ratio between the flux as would be measured by a 22\arcsec\ Gaussian beam to the total flux in the map ($S_{CO,obs}/S_{tot}$).  We find that a single central observation recovers most $(>60\%)$ of the CO line flux in galaxies with $D_{25}<40\arcsec$.  Results are shown in Figure \ref{figapcor}a.  Based on the best fit to these data, we apply the following aperture correction:
\begin{equation}
S_{CO,cor} =  S_{CO,obs} / (1.094 - 0.008 D_{25} + 2.0\times10^{-5} D_{25}^2)
\label{Icor1}
\end{equation}
where $S_{CO,obs}$ is the observed flux in the central pointing and $S_{CO,cor}$ the extrapolated total flux.

For galaxies with optical diameters larger than 40\arcsec, however, there is a significant scatter in the $S_{CO,obs}/S_{tot}$ ratio, and additional information is required to recover the total CO line flux.  In the right panel of Figure \ref{figapcor}, we show that with a single offset pointing at three-quarter beam from the central position ($0.75\times$22\arcsec) along the major axis, the total flux can be recovered with much better accuracy\footnote{During the pilot observations of 2009 June, we took offset pointings one full beam from the central position.  These galaxies are identified clearly by asterisks in Col. 6 of Tab. \ref{COtab}, and aperture corrections are performed using a version of Eq. \ref{Icor2} appropriate for these larger offsets.}.  An offset pointing at a full beam also does well, but the mean ratio $S_{off}/S_{CO,obs}$ then drops from 33\% to 15\%.  We adopted the three-quarter beam offset as a compromise between the requirements for independent flux measures and a modest fraction of our total observing time going into off-center pointings.  

Our requirement to perform an offset pointing is that a galaxy has (1) a large angular size ($D_{25}>40$\arcsec) and (2) a bright CO line in the central pointing, such that a detection in the offset pointing can be made with an hour of integration or less.  For these galaxies, the flux in the central pointing is corrected based on the ratio between the flux in the offset pointing to that in the center ($f_{off}\equiv S_{off}/S_{CO,obs}$), based on the best fit to data in Figure \ref{figapcor}b:
\begin{equation}
S_{CO,cor}=S_{CO,obs} / (1.587 - 3.361 f_{off} + 2.107 f_{off}^2).
\label{Icor2}
\end{equation}
If a galaxy is larger than 40\arcsec\ but the CO line in the central pointing is weak or undetected, then the measured central flux is corrected according to Eq. \ref{Icor1}.

So far, we have performed offset pointings for \noff galaxies, that met the requirements listed above (see Figs. \ref{spectra_off} and \ref{spectra_off2}).  We used Eq. \ref{Icor2} and Fig. \ref{figapcor}b to infer the fraction of the total flux that was measured in the central pointing ($S_{CO,obs}/S_{tot}$), and we show in Fig. \ref{figapcor}a how this ratio depends on $D_{25}$ and how it compares to the range of values predicted by our simulated observations.

\subsection{$M_{H2}$ and associated error budget}

After correcting the CO(1-0) line fluxes for aperture effects using either Eq. \ref{Icor1} or Eq. \ref{Icor2}, whichever case applies to each galaxy, we compute the total CO luminosities following \citet{solomon97}:
\begin{equation}
L'_{CO}=3.25\times10^7 S_{CO,cor} \nu_{obs}^{-2} D_{L}^2 (1+z)^{-3},
\label{Ico}
\end{equation}
where $S_{CO,cor}$ in units of Jy \kms is the integrated line flux\footnote{Calculated from antenna temperature units using the conversion $S/T_a^{\ast}=6.0 {\rm Jy/K}$, specific for the IRAM 30m at our observing frequency of 111GHz.}, $\nu_{obs}$ is the observed frequency of the CO(1-0) line in GHz,  $D_L$ is the luminosity distance in units of Mpc, and $L'_{CO}$ is the CO luminosity in [K \kms\ pc$^2$]. 

The total molecular hydrogen masses are then calculated as $M_{H2}=L'_{CO}\alpha_{CO}$.  We adopt a constant Galactic conversion factor of $\alpha_{CO}=3.2$ M$_{\odot}$(K \kms\ pc$^2)^{-1}$, which does not include a correction for the presence of Helium.  Our choice of $\alpha_{CO}$ is roughly the mean of values estimated in the Milky Way and in nearby galaxies \citep[e.g.][]{strong96,dame01,blitz07,draine07,heyer09,abdo10}.  The virial method used to measure $\alpha_{CO}$ has been validated by other independent techniques such as $\gamma-$ray observations, and shown to also hold for the ensemble average of the virialized clouds of entire galaxies, as long as the factor $n({\rm H}_2)^{0.5}/T$ is constant throughout the galaxy and the CO line is optically thick \citep{dickman86,tacconi10}.

This constant value for $\alpha_{CO}$ has been shown to hold for galaxies in the Local Group, when the metallicity goes from solar down to SMC values \citep{bolatto08}.  It may be that the conversion factor is instead a function of a parameter such as gas surface density or metallicity \citep{tacconi08,obreschkow09}.  Based on the well-known metallicity-luminosity relation, \citet{boselli02} for example proposed a luminosity-dependent conversion factor.  However, such a prescription has yet to be observationally or theoretically validated.  Furthermore, it has since been shown that the mass-metallicity relation is even more fundamental \citep{tremonti04}.  At stellar masses above $\sim 10^{10.5}$\msun, the relation flattens out, and therefore the metallicity of our COLD GASS galaxies is expected to be $\sim$solar with little variations across the sample.  This expectation is confirmed by our long-slit spectroscopy measurements \citep{moran10,COLDGASS2}. Based on this fact, adopting a constant Galactic conversion factor $\alpha_{CO}=3.2$ M$_{\odot}$(K \kms\ pc$^2)^{-1}$ is the simplest yet most justified assumption we can make at this point. We however consider the variations in $\alpha_{CO}$ by a factor of $\sim$2 at fixed metallicity \citep{bolatto08} as the systematic uncertainty on our values of $M_{H_2}$.

We calculate the formal measurement error on the observed line flux, $S_{CO,obs}$, as:
\begin{equation}
\epsilon_{obs}=\frac{\sigma_{rms} W50_{CO}}{\sqrt{W50_{CO} \Delta w_{ch}^{-1}}}, 
\label{errmes}
\end{equation}
where $\sigma_{rms}$ is the rms noise per spectral channel of width $\Delta w_{ch}=21.57$\kms, and $W50_{CO}$ is the FWHM of the CO(1-0) line.  This definition takes into account that for a given total flux, the $S/N$ per channel is largest when the emission line is narrower.  Considering all detections, the mean fractional error ($\epsilon_{obs}/S_{CO,obs}$) is $11\%$.

Other contributions to the error budget on $M_{H2}$ include a flux calibration error ($10\%$ at a wavelength of 3mm, under average atmospheric conditions) and the uncertainty on the aperture correction (which we estimate to be $15\%$ based on Fig.\ref{figapcor} for the median galaxy in our survey).  The average pointing rms error is $\Delta_{point}=2\arcsec$, and also contributes to the uncertainty on the measured flux.   Using the resolved CO(1-0) maps of \citet{kuno07}, we simulate the observation of COLD GASS galaxies both with a beam perfectly centered on the object, and with a beam offset by $\Delta_{point}$ from the center.  We find that a $2\arcsec$ positional error generates a 2.1\% uncertainty on the measured line flux.   Given that the redshift error is negligible compared to these other contributions, the fractional error on $L'_{CO}$ is obtained by adding in quadrature all error contributions to $S_{CO,cor}$.  We therefore find a mean error of $<\epsilon_{Lco}>\simeq20\%$.  The random measurement error on our quoted values of $M_{H2}$ is then $\sim20\%$, to which we add the systematic error coming from the uncertainty in the value of $\alpha_{CO}$ bringing the total error on $\log M_{H_2}$ to 0.3 dex.

\subsection{Stacking}
\label{stacking}
To extract information from the CO non-detections, we also perform a stacking analysis.  We stack the spectra after converting them in ``gas fraction" units following \citet{fabello10}, who provide an extensive discussion of the merits of this technique.  We convert each spectrum $S$ into a scaled spectrum $S_{gf}$:
\begin{equation}
S_{gf}=\frac{S D_L^2}{M_{\ast} (1+z)^3},
\end{equation}
where the factor $D_L^2 (1+z)^{-3}$ is a consequence of Eq. \ref{Ico}.  This gives us scaled spectra in units of [mJy Mpc$^2$ \msun$^{-1}$].  

For any subsample of $N$ galaxies with CO non-detection, we calculate a mean stacked spectrum:
\begin{equation}
S_{stack}=\frac{\sum_{i=1}^N w_i S_{gf,i}}{\sum_{i=1}^N w_i},
\end{equation}
 where $w$ is the weight given to each galaxy in the stack, based on its stellar mass, in order to compensate for the fact that the $\log M_{\ast}$ distribution of the COLD GASS sample is by design flatter than in a complete volume-limited sample (Figure \ref{distribs}a).

To obtain $<M_{H2}/M_{\ast}>$, the mean molecular gas mass fraction for the family of $N$ galaxies stacked, we measure a line flux or set an upper limit as described in \S \ref{linemes}, and multiply by the constant factors $3.25\times10^7 \nu_{obs}^{-2} \alpha_{CO}$ to obtain the dimensionless mass fraction.

\subsection{Catalog presentation}
\label{catalog}

In table \ref{COtab},  we summarize the results of the IRAM observations of the first \ntot\ COLD GASS galaxies (\ndet\ CO detections and \nondet\ non-detections).   The contents of the catalog are as follows: \\
\\
Column (1): GASS ID.  Galaxies are in the same order as in Table \ref{params} to ease cross-referencing. \\
\\
Column (2): rms noise per channel in mK, after binning the spectra to a resolution of $\Delta w_{ch}=21$ \kms.  \\
\\
Column (3): Signal-to-noise ratio of the detected CO line, calculated as $S_{CO,obs}/\epsilon_{obs}$, where $\epsilon_{obs}$ is calculated according to Eq. \ref{errmes}.  In our analysis, we consider sources with $S/N>5$ as secure, and the detections with $S/N<5$ as tentative.\\
\\
Column (4): Integrated CO line flux in Jy \kms. The measured flux in antenna temperature units is converted to these units using the ratio $S/T_a^{\ast}=6.0$ Jy/K for the IRAM 30m telescope at our observing frequency of 111GHz.\\
\\
Column (5): Corrected total line flux in Jy \kms, computed following Equations \ref{Icor1} and \ref{Icor2}. \\
\\
Column (6): When available, the measured ratio between the flux in the offset pointing
to that in the central pointing, as described 
in \S \ref{offsets}. \\
\\
Column (7): Total molecular hydrogen gas mass.  We adopt a constant Galactic conversion factor of $\alpha_{CO}=3.2$ M$_{\odot}$(K \kms\ pc$^2)^{-1}$, which does not include a correction for the presence of Helium.  The numbers quoted are either measured masses in the case of detections, or $5\sigma$ upper limits for the non-detections (see Column 9).  \\
\\
Column (8): Molecular gas mass fraction, $f_{H2} \equiv M_{H2}/M_{\ast}$, and upper limits for non-detections. \\
\\
Column (9): CO emission line flag, set to 1 for detections and 2 for non-detections.

\begin{table*}
\begin{minipage}{130mm}
\caption{Molecular gas masses and CO(1-0) parameters for the COLD GASS galaxies$^a$}
\label{COtab}
\begin{tabular}{cccccccccc}
\hline
GASS ID & $\sigma$ & $S/N$ & $S_{CO,obs}$ & $S_{CO,cor}$ &
$f_{off}$ & $M_{H2}$  & $M_{H2}/M_{\ast}$ & Flag \\
 &[mK] &  & $[\rm{Jy~km~s}^{-1}]$ & $[\rm{Jy~km~s}^{-1}]$ &
 & $[\log M_{\odot}]$ & & \\
\hline
 11956 & 1.07 &  2.35 &  1.16 &  1.25    & \nodata &  8.46 & 0.024 &    1 \\
 12025 & 1.06          &  \nodata  & \nodata & \nodata & \nodata &  8.78 & 0.009 &    2 \\
 12002 & 1.18          &  \nodata  & \nodata & \nodata & \nodata &  8.79 & 0.021 &    2 \\
 11989 & 1.07          &  \nodata  & \nodata & \nodata & \nodata &  8.79 & 0.013 &    2 \\
 27167 & 1.17          &  \nodata  & \nodata & \nodata & \nodata &  8.74 & 0.023 &    2 \\
  3189 & 1.24 &  6.69 &  3.19 &  3.88    & \nodata &  8.93 & 0.076 &    1 \\
  3261 & 1.96 &  8.57 &  4.27 &  4.62    & \nodata &  8.98 & 0.080 &    1 \\
  3318 & 1.03          &  \nodata  & \nodata & \nodata & \nodata &  8.81 & 0.019 &    2 \\
  3439 & 1.03          &  \nodata  & \nodata & \nodata & \nodata &  8.79 & 0.028 &    2 \\
  3465 & 1.17 &  4.28 &  2.89 &  3.28    & \nodata &  8.62 & 0.027 &    1 \\
  3645 & 1.08          &  \nodata  & \nodata & \nodata & \nodata &  8.62 & 0.020 &    2 \\
  3509 & 1.15 &  7.24 &  5.02 &  5.80    & \nodata &  9.30 & 0.031 &    1 \\
  3519 & 1.53 &  7.72 &  5.51 &  6.52    & \nodata &  9.24 & 0.032 &    1 \\
  3505 & 1.30          &  \nodata  & \nodata & \nodata & \nodata &  8.93 & 0.052 &    2 \\
  3504 & 1.53 &  9.19 &  2.83 &  3.44    & \nodata &  8.87 & 0.051 &    1 \\
\hline
\end{tabular}
$^a$ Table 1 is published in its entirety in the electronic edition of the journal.  A portion is shown here as example of its format and contents.
\end{minipage}
\end{table*}


\section{Molecular gas fraction scaling relations}
\label{scalrel}

\subsection{Molecular gas fraction and global galaxy parameters}
\label{fH2_s1}

\begin{figure*}
\begin{minipage}{165mm}
\includegraphics[width=150mm]{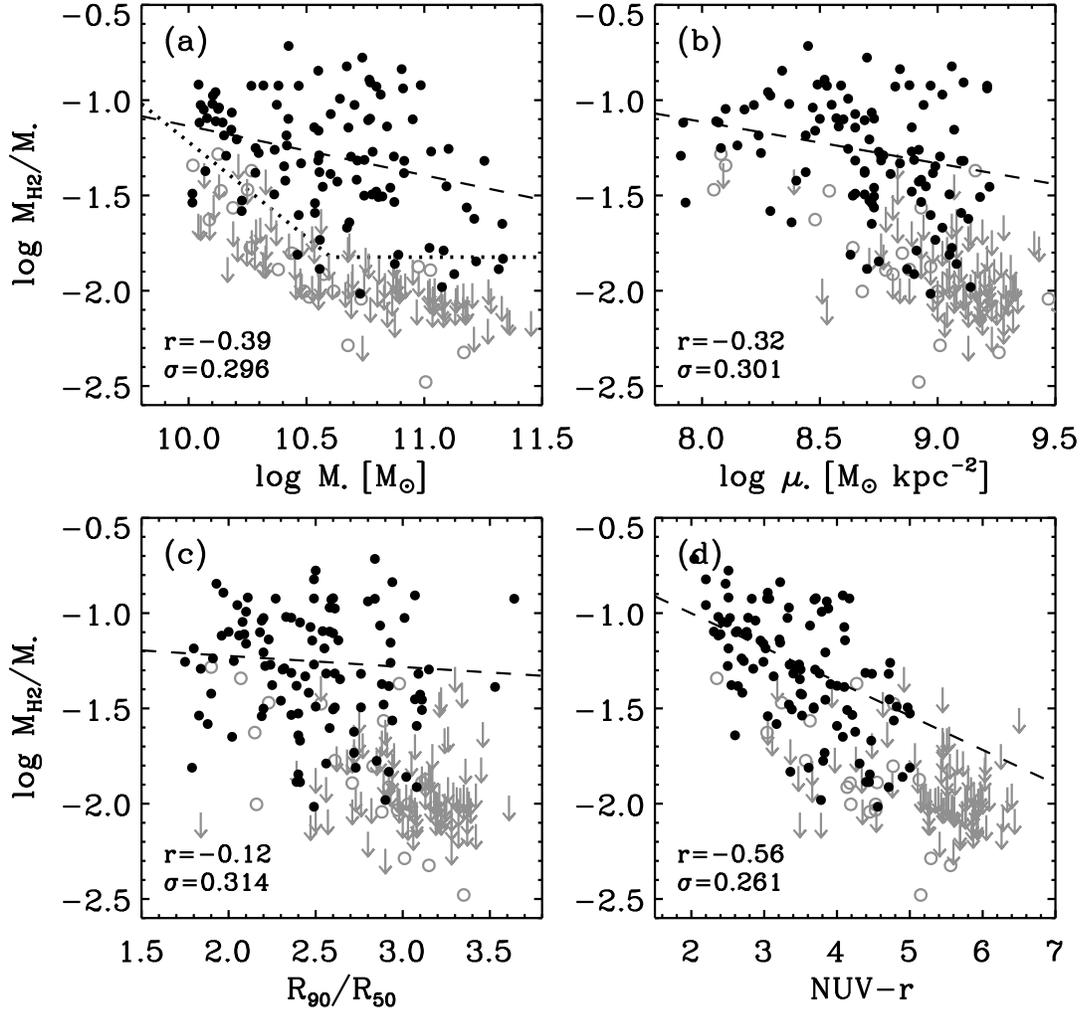}
\caption{Molecular gas mass fraction ($f_{H2}\equiv M_{H2}/M_{\ast}$) as a function of {\it (a)} stellar mass, {\it (b)} stellar mass surface density, {\it (c)} concentration index $R_{90}/R_{50}$, and{\it (d)} NUV$-r$ colour .  Detections in CO are shown as circles (filled: $S/N>5$, open: $S/N<5$), 
and upper limits obtained for non-detections as arrows.  In panel {\it (a)}, the dotted line represents the detection limit of the survey, which is a function of \mstar\ at the low mass end due to the hard limit of 1.1mK for the observed rms (see Section \ref{obsstrat} for details). In each panel, we give the value of the Pearson correlation coefficient of the relation ($r$), and the standard deviation $\sigma$ about the best fit linear relation to all CO detections (dashed line). \label{fH2plots}}
\end{minipage}
\end{figure*}

We first investigate how the molecular gas mass fraction\footnote{We warn the reader that our definition of $f_{H2}$ differs from some previous studies, where $f_{H2}$ is also defined as $M_{H2}/M_{dyn}$ or $M_{H2}/(M_{\ast}+M_{HI}+M_{H2})$.} ($f_{H2}\equiv M_{H2}/M_{\ast}$) varies across the COLD GASS sample.   The values of $f_{H2}$ range from 0.009 (G7286), at the very edge of the survey's detection limit, up to 0.20 for the most $H_2-$rich galaxy in the sample observed so far (G41969).  Overall, the molecular gas contents are more modest, averaging at $<f_{H2}>=0.066\pm0.039$ for the CO detections, and at $0.043\pm0.022$ when non-detections are included in the mean at the value of their upper limits.   These sample averages are measured by weighting each galaxy differently according to its stellar mass, in order to compensate for the flat $\log M_{\ast}$ distribution (see Section \ref{fH2_s2} for more details).

In Figure \ref{fH2plots}, we investigate how the molecular gas mass fraction depends on stellar mass, stellar mass surface density, concentration index and \nuvr\ colour.   For all four relations, we quantify the strength of the dependence of $f_{H2}$ on the x-axis parameter in two ways.  Firstly, we compute for all secure detections the Pearson correlation coefficient of the relation, $r$. Secondly we perform an ordinary least squares linear regression of $f_{H2}$ on the x-axis parameter for the secure detections, and report the standard deviation of the residuals about this best fitting relation, $\sigma$, in $\log M_{H2}/M_{\ast}$ units. 

There is a weak dependency of $f_{H2}$ on stellar mass, $M_{\ast}$, as seen in Fig. \ref{fH2plots}a, with a correlation coefficient $r=-0.40$, and a scatter of $\sigma=0.292$ dex, but some of this correlation is caused by the detection limit.   The strength of the correlation of $f_{H2}$ on stellar mass surface density, $\mu_{\ast}$, is similar with $r=-0.33$ and $\sigma=0.301$ dex (Fig. \ref{fH2plots}b).  There is however a striking difference in the behaviour of the non-detections; while the detection rate is roughly constant as a function of \mstar ($\sim50\%$, see Fig. \ref{detfrac}a), it is a strong function of \must.  There is a surface density threshold of $\mu_{\ast}\simeq10^{8.7}$ \msun\ kpc$^{-2}$, below which the detection rate is almost $100\%$ and above which it quickly drops.  A similar threshold is observed by \citet{GASS1} in the relation between the atomic gas mass fraction ($f_{HI}\equiv M_{HI}/M_{\ast}$) and \must\ (see \S \ref{compHI} for more details).

To test the impact of galaxy morphology on the molecular mass fraction, we plot in Fig. \ref{fH2plots}c $f_{H2}$ as a function of the concentration index, defined as $C\equiv R_{90}/R_{50}$ where $R_{90}$ and $R_{50}$ are the radii enclosing 90\% and 50\% of the $r-$band flux, respectively.  The concentration index is a good proxy from the bulge-to-total ratio, as would be recovered by full two-dimentional bulge/disk decompositions \citep{weinmann09}.  The galaxies with a CO line detection show no correlation, with a correlation coefficient of $-0.14$, and a scatter about the best-fit relation (0.317 dex) as large as the scatter in $f_{H2}$ itself (also 0.317 dex), suggesting that the presence of a bulge does not affect strongly how much of the mass in these galaxies is in the form of molecular gas.  A similar result was found by \citet{GASS1} and \citet{fabello10} for the HI contents of similar galaxies. The only effect seen is again in the detection fraction, with the success rate of measuring the CO line dropping sharply from $100\%$ for the most disk-dominated systems, to $0\%$ for the most bulge-dominated ones. 

\begin{figure}
\includegraphics[width=84mm]{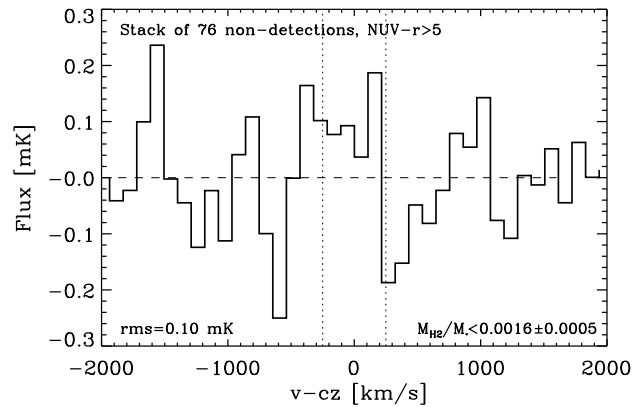}
\caption{Stacked spectrum of all galaxies with \nuvr$>5.0$.  Individually, the galaxies included are all non-detections in CO, and even their stacked average does not yield a detection, setting instead a stringent upper limit of $<M_{H2}/M_{\ast}>=0.0016\pm0.0005$. The vertical dotted line show an expected line width of $\sim 500$\kms\  for such a stacked signal.  \label{figstack}}
\end{figure}

The only parameter upon which the measured molecular mass fractions does depend significantly is colour ($r=-0.57$, $\sigma=0.260$ dex, see Fig. \ref{fH2plots}d).  Since \nuvr\ colour is a proxy for specific star formation rate, 
this dependency of $f_{H2}$ is expected, because star formation and
molecular gas are known to be strongly correlated.  
It is interesting to note that not a single red sequence 
galaxy (\nuvr\ $>5.0$) has a measurable molecular 
gas component ($f_{H2}<0.015$ in all cases).   To test this further, we stack all non-detections with \nuvr$>5$ using the technique described in \S \ref{stacking}.   The result is shown in Figure \ref{figstack}.  Even in the stack, the red galaxies lead to a non-detection of the CO line, thus setting an even more restrictive upper limit on the molecular gas mass fraction of $0.0016\pm0.0005$, which at their median stellar mass corresponds to an upper limit on the molecular gas mass of $\sim1.1\times10^{8} M_{\odot}$.  Molecular gas has been detected in early-type galaxies \citep[e.g. in the SAURON sample,][]{combes07}, but at levels even below this limit. Our results are therefore not in contradiction with these studies.

\subsection{Mean scaling relations}
\label{fH2_s2}

\begin{figure*}
\begin{minipage}{165mm}
\includegraphics[width=150mm]{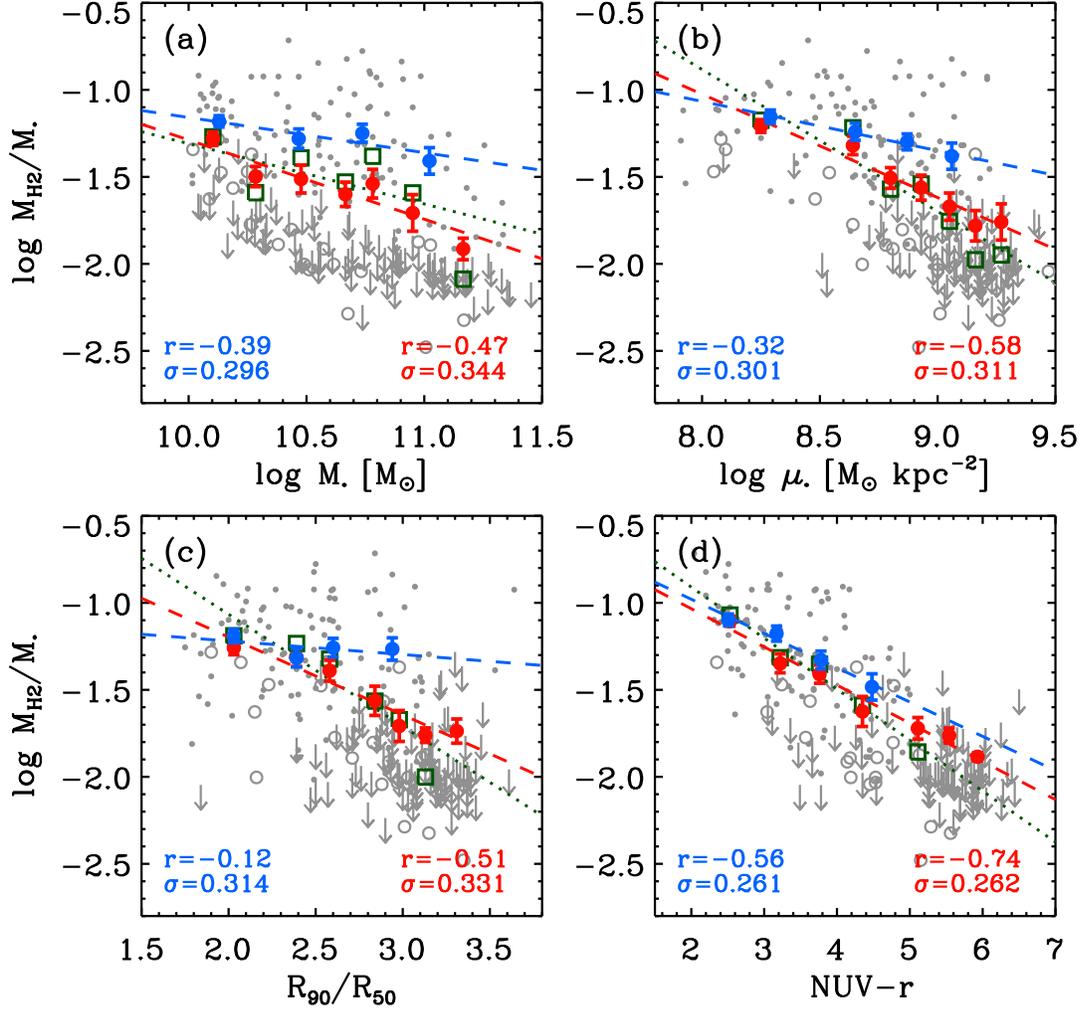}
\caption{Molecular gas mass fraction scaling relations. The weighted mean values of  $\log f_{H2}$ in equally populated bins are plotted as a function of {\it (a)} stellar mass, {\it (b)} stellar mass surface density, {\it (c)} concentration index $R_{90}/R_{50}$, and{\it (d)} NUV$-r$ colour.  The blue circles include only CO detections.  The red circles and green squares include both detections and non-detections, but differ in the way the latter are treated.  We assign to the non-detections either the value set by the upper limit (red circles), or simply set them to zero (green squares).  The correlation coefficients ($r$) and scatter around the best-fit relation are given for detections only (lower left corners) and for the ``non-detections as upper limits" case (lower right corners). The gray symbols show the data for all COLD GASS galaxies within this data release (as presented in Fig. \ref{fH2plots}).  \label{fH2scalrel}}
\end{minipage}
\end{figure*}

After having looked separately at the properties of CO detections and non-detections as a function of several global galaxy parameters (Fig. \ref{fH2plots}), we combine all measurements into mean molecular gas mass fraction scaling relations.   

As described in \S \ref{sampleselect}, the COLD GASS sample has been generated to have a stellar mass distribution that is flatter than it is in a purely volume-limited sample (see Fig. \ref{distribs}).  When building the scaling relations, we correct for this by weighing each point according to its stellar mass.  Following \citet{GASS1}, the galaxies are placed in bins of stellar mass of width 0.2 dex and assigned as a weight the ratio between the total number of galaxies in the unbiased volume-limited parent sample and the total number of COLD GASS galaxies within that same mass bin.  In other words, low mass galaxies are given a higher weight in the computation of the mean scaling relations, because these galaxies are underrepresented in the COLD GASS sample compared to a volume-limited sample. 

\begin{table*}
\begin{minipage}{175mm}
\caption{Mean molecular gas fraction scaling relations$^a$}
\label{ALLtab}
\begin{tabular}{cccccccccc}
\hline
 & & \multicolumn{2}{c}{(1) Detections only} & & \multicolumn{2}{c}{(2) All, non-detections=limit} & & \multicolumn{2}{c}{(3) All, non-detections=0}\\
\cline{3-4} \cline{6-7} \cline{9-10} \noalign{\smallskip}
$x$ & $x_0$ & $m$ & $b$ & & $m$ & $b$ & & $m$ & $b$ \\
\hline
          $\log M_{\ast}$ & 10.70 & $-0.202\pm0.040$ & $-1.300\pm0.592$ & &
  $-0.455\pm0.069$ & $-1.607\pm1.034$ & &
  $-0.346\pm0.203$ & $-1.552\pm3.038$ \\
        $\log \mu_{\ast}$ &  8.70 & $-0.283\pm0.019$ & $-1.265\pm0.237$ & &
  $-0.595\pm0.055$ & $-1.441\pm0.682$ & &
  $-0.816\pm0.127$ & $-1.454\pm1.587$ \\
          $R_{90}/R_{50}$ &  2.50 & $-0.078\pm0.048$ & $-1.258\pm0.176$ & &
  $-0.447\pm0.049$ & $-1.420\pm0.185$ & &
  $-0.644\pm0.145$ & $-1.388\pm0.529$ \\
                  NUV$-r$ &  3.50 & $-0.197\pm0.012$ & $-1.275\pm0.066$ & &
  $-0.219\pm0.013$ & $-1.363\pm0.077$ & &
  $-0.293\pm0.013$ & $-1.349\pm0.069$ \\
\hline
\end{tabular}
$^a$ The relations are parametrized as $\log t_{dep}({\rm H}_2)[{\rm yr}^{-1}]=m(x-x_0)+b$.
\end{minipage}
\end{table*}

Each scaling relation is computed and plotted using three different subsamples: (1) only the galaxies with CO detections are considered, (2) both the detections and non-detections are used, and the upper limit on $f_{H2}$ is used for the non-detections, and (3) both the detections and non-detections are used, but this time a value of $f_{H2}=0.0$ is assigned to the non-detections.   In all cases, the weighted mean of $\log f_{H2}$ is then calculated in equally populated bins of either \mstar, \must, concentration index or \nuvr\ colour, and the relations are fitted, weighting galaxies according to their stellar mass.  The resulting mean relations are plotted in Figure \ref{fH2scalrel}, with the error bars representing the uncertainty on the position of $<\log f_{H2}>$ in equally populated bins, as determined by bootstrapping: the error is the standard deviation in the value of  $<\log f_{H2}>$ for 1000 resamples of the original data in each bin, with repetitions.  The best-fit linear relations are also plotted and summarized in Table \ref{ALLtab}. Note that in case (3), because of the null values of $f_{H2}$ for the non-detections, we cannot directly average and fit the values of $\log f_{H2}$ as we do for cases (1) and (2).  Instead, we measure the logarithm of the mean value of $f_{H2}$ in each bin, and fit these average values.  This is the relation plotted in Figure \ref{fH2scalrel} and given in Table \ref{ALLtab} for case (3).

The choice between setting the non-detections to their upper limits or to a constant value of zero only significantly affects the scaling relations at large values of \must, $C$ and \nuvr\, where few galaxies have a detected CO line.  Both sets of scaling relations are included in Table \ref{ALLtab}, labeled as $<\log M_{H2}/M_{\odot}>_{lim}$ and $<\log M_{H2}/M_{\odot}>_{0}$ for the upper limit and zero value cases, respectively.

The mean molecular gas mass fraction is a roughly constant function of stellar mass, both for detections alone and when including non-detections, as shown in Fig. \ref{fH2scalrel}a.  This is a consequence of the flat detection rate of the CO line as a function of \mstar.  Adding non-detections turns the mostly-flat relations between $f_{H2}$ and both \must\ and $C$ into monotonically decreasing functions.  The strongest correlation is still with \nuvr\ colour, both before and after including non-detections.  Because \nuvr\ colour is a proxy for specific star formation rate, its correlation with $f_{H2}$ is not surprising, given e.g. the Kennicutt-Schmidt relation.   The correlation of $f_{H2}$ with \nuvr\ could therefore be seen as a consequence rather than a cause. 

For the other parameters describing the underlying properties of the galaxies, independently of the current star formation rate (\mstar, \must, concentration index),  there is very little dependence of the measured values of $f_{H2}$.   In \S \ref{fH2_s1}, we showed that no matter their stellar mass, about half of our sample has detectable CO line emission, which converts into a molecular mass fraction of $\sim 6\%$ across the stellar mass interval sampled.  On the other hand, while the molecular mass fraction of $detected$ galaxies is mostly independent of \must, the detection fraction is a strong function of that quantity.  These observations trace a picture where: (1) the conditions required for the formation/consumption of molecular gas are a strong function of \must\ (and concentration index) but not of \mstar, but (2) when these conditions are met, a roughly constant fraction of the stellar mass is found in the form of molecular gas.


\section{The relationship between HI and $H_2$}
\label{compHI}

\begin{figure}
\includegraphics[width=84mm]{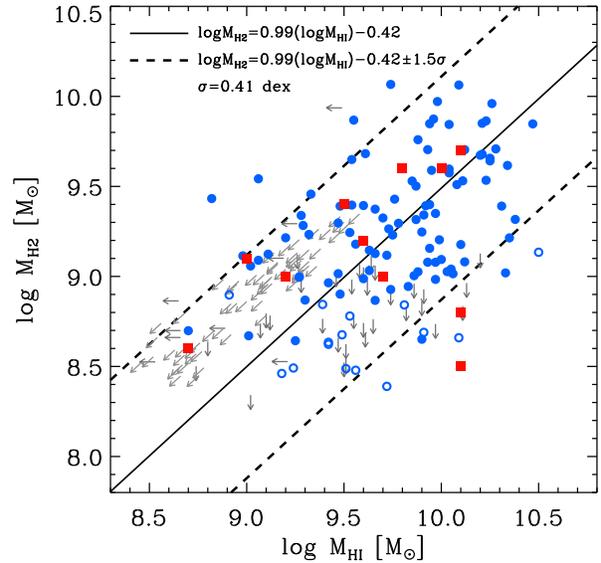}
\caption{Comparison between atomic and molecular hydrogen gas masses.  COLD GASS galaxies with detections of both the HI and CO lines are plotted as filled and open blue circles, for secure and tentative CO detections, respectively.  The arrows shows limits in the cases of non-detecion of either or both the HI and CO lines. The best bisector linear fit to the detections is show as a solid line, and the $\pm 1.5 \sigma$ region around this fit marked with dashed lines. For comparison, we overplot as filled squares the integrated measurements for the HERACLES galaxies with $M_{\ast}>10^{10}M_{\odot}$, taken from \citet{leroy08}. \label{HIH2}}
\end{figure}

Under the assumption that molecular gas forms out of lower density clouds of atomic gas, one might  na\"{i}vely expect a tight correlation between $M_{HI}$ and $M_{H2}$.  The actual situation is, however, quite different, as seen in Figure \ref{HIH2}.  Within the subsample of galaxies detected both in HI and CO, the fraction $M_{H2}/M_{HI}$ varies greatly, from 0.037 (G13775) up to 4.09 (G38462).  For galaxies with both CO and HI-detections, the correlation coefficient between $\log M_{HI}$ and $\log M_{H2}$ is $r=0.37$, indicating that the two quantities are only weakly correlated. A bisector linear fit to the same subsample reveals that on average, $M_{H2}$ is 0.295 times the value of $M_{HI}$, with a large scatter of 0.41 dex in $\log(M_{H2}/M_{HI})$ (see Fig. \ref{HIH2}). 

\begin{figure*}
\begin{minipage}{165mm}
\includegraphics[width=150mm]{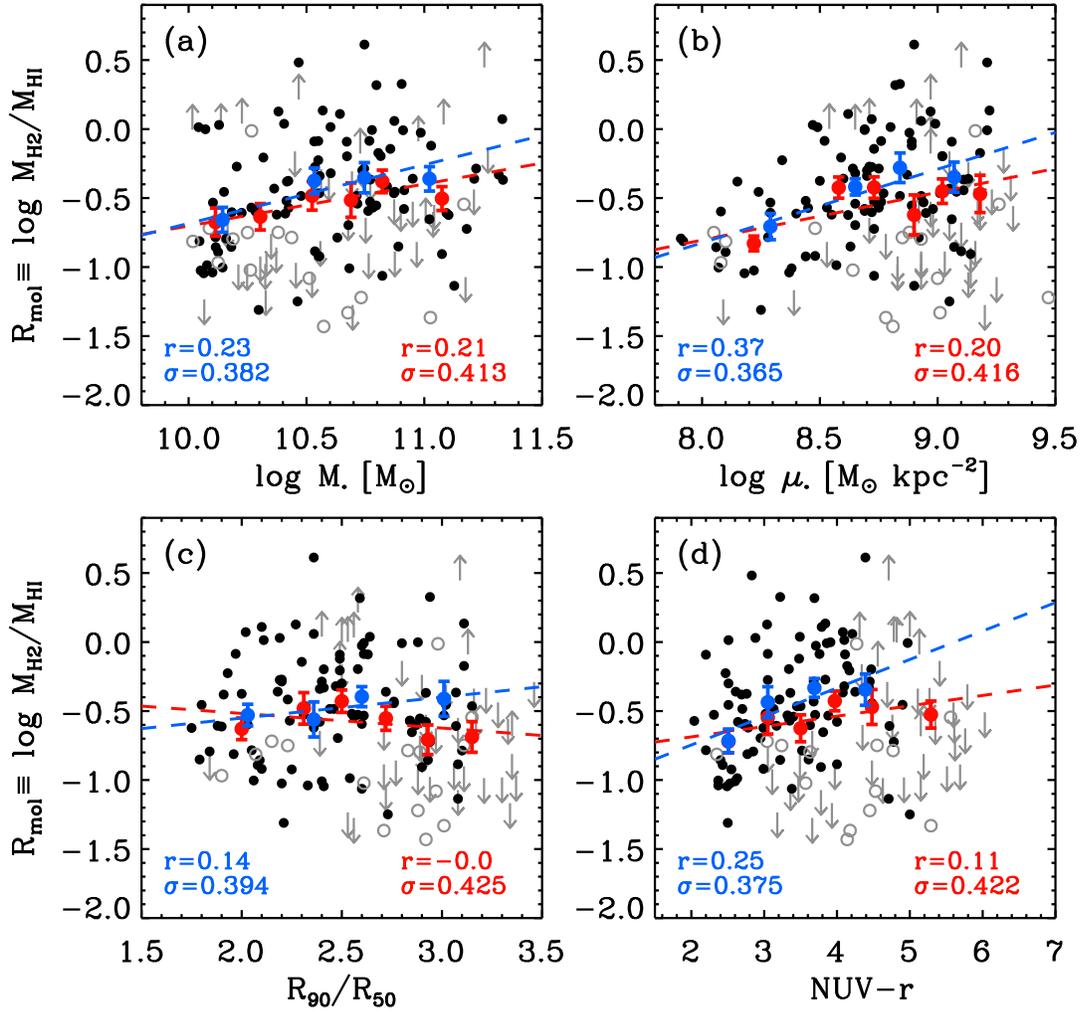}
\caption{Molecular to atomic gas mass ratio as a function of {\it (a)} stellar mass, {\it (b)} stellar mass surface density, {\it (c)} concentration index $R_{90}/R_{50}$, and{\it (d)} NUV$-r$ colour .  Detections in CO are shown as circles (filled: $S/N>5$, open: $S/N<5$), non-detections as upper limits.  Non-detections in HI but with a CO detection are shown as lower limits.  The mean scaling relations are also overplotted, with symbols as described in Fig. \ref{fH2scalrel}. \label{MH2MHI}}
\end{minipage}
\end{figure*}

To further investigate the relationship between $M_{H2}$ and $M_{HI}$, we look into how the ratio between these two quantities varies as a function of different physical parameters.  We define the molecular fraction as the ratio between the molecular hydrogen gas mass and the atomic gas mass of the system:
\begin{equation}
R_{mol}=\log \left( \frac{M_{H2}}{M_{HI}} \right). 
\label{fmol}
\end{equation}
In Figure \ref{MH2MHI}, we plot \rmol\ as a function of the same four global parameters as in the previous figures. Galaxies with non-detections in both HI and $H_2$ are not plotted, as $R_{mol}$ is completely unconstrained in these cases.  he best fit linear relations, taking into account the weights correcting for the flat $\log M_{\ast}$ distribution, are measured in two ways: (1) including only the secure detections in HI and \hmol, and (2) including the non-detections in either of these quantities as lower or upper limits, respectively.  In Figure \ref{MH2MHI}, we also report the Pearson correlation coefficients and the scatter around these relations.  The best-fit scaling relations are given the form $R_{mol}=m(x-x_0)+b$, with the parameters presented in Table \ref{Rmoltab}.

\begin{table*}
\begin{minipage}{125mm}
\begin{center}
\caption{Molecular-to-atomic gas mass ratio relations$^a$}
\label{Rmoltab}
\begin{tabular}{ccccccc}
\hline
 & & \multicolumn{2}{c}{(1) Detections only} & & \multicolumn{2}{c}{(2) All, non-detections=limit}\\
\cline{3-4} \cline{6-7} \noalign{\smallskip}
$x$ & $x_0$ & $m$ & $b$ & & $m$ & $b$ \\
\hline
          $\log M_{\ast}$ & 10.70 & $ 0.425\pm0.097$ & $-0.387\pm1.464$ & &
  $ 0.303\pm0.084$ & $-0.489\pm1.253$ \\
        $\log \mu_{\ast}$ &  8.70 & $ 0.533\pm0.128$ & $-0.451\pm1.562$ & &
  $ 0.346\pm0.142$ & $-0.562\pm1.772$ \\
          $R_{90}/R_{50}$ &  2.50 & $ 0.152\pm0.031$ & $-0.475\pm0.113$ & &
  $-0.106\pm0.098$ & $-0.571\pm0.367$ \\
                  NUV$-r$ &  3.50 & $ 0.206\pm0.057$ & $-0.436\pm0.289$ & &
  $ 0.075\pm0.032$ & $-0.575\pm0.169$ \\
\hline
\end{tabular}
\end{center}
$^a$ The relations are parametrized as $\log M_{H_2}/M_{HI}=m(x-x_0)+b$.
\end{minipage}
\end{table*}

The first panel shows that \rmol\ is a very weakly increasing function of \mstar\ ($r=0.23$).  Both with and without the non-detections included, the slope of the relation is positive at the 3$\sigma$ significance level (see Table \ref{Rmoltab}).  As demonstrated previously, $f_{H2}$ does not appear to depend significantly on \mstar (Fig. \ref{fH2plots}a), while $f_{HI}$ is a fairly strongly declining function of \mstar\ \citep{GASS1}.  The combination of these two trends produces the weak increase in $R_{mol}$ as a function of stellar mass.

\citet{GASS1} also reported a strong anti-correlation of $f_{HI}$ with the stellar mass surface density, \must.   For galaxies with CO detections, the values of $f_{H2}$  show a similar, but considerably weaker trend as a function of \must.  In their resolved study, \citet{leroy08} show that the molecular-to-atomic ratio is a strong function of local properties within the disks of spiral galaxies. In particular, they find a dependence on stellar mass surface density, with the molecular fraction steadily increasing from surface mass densities of $10^{7.5}$ to $10^9$ \msun\ kpc$^{-2}$. Our molecular fractions are smaller because our measurements are integrated over entire galaxies, but the same qualitative trend is observed for our global measurements.

Even though $f_{HI}$ and $f_{H2}$ show different dependencies on \must, the fraction of galaxies with non-detections as a function of \must\ in the HI and CO samples exhibits very similar behaviour.   As shown in Figure \ref{fH2plots}b and in \citet{GASS1},  there is a critical mass surface density of $\mu_{\ast}=10^{8.7}$ \msun\ kpc$^{-2}$ below which all galaxies have a sizeable HI and $H_2$ component, and above which cold gas seems to have mostly disappeared.  There are similar thresholds in concentration index and \nuvr\ colour (see e.g. our detection fractions in Fig. \ref{detfrac}).  However, while the mean value of $f_{HI}$ never falls below $\sim2\%$, even at high \must\ or \nuvr\ colour \citep{fabello10}, $f_{H2}$ drops sharply below that level, as shown in Fig. \ref{fH2scalrel} and evidenced by the results of our stacking experiment (Fig. \ref{figstack}).  Furthermore, even though \citet{GASS1} found some red sequence galaxies with a surprisingly large HI component, none of these galaxies have a sizeable molecular gas mass; none of the galaxies with \nuvr$>5$ are securely detected in CO.  It therefore seems that above our empirical thresholds in \must, concentration index and \nuvr\ colour, an increasing fraction of galaxies with any form of cold gas at all {\em appears to be dominated by atomic gas}.  This result is also striking in  Fig. \ref{MH2MHI}c, where we see the same population of atomic-gas dominated galaxies (i.e. the upper limits in $R_{mol}$) almost exclusively at $C>2.6$.   We note that this critical value of concentration index (C=2.6) corresponds to the observed transition between late- and early-type galaxies \citep[e.g.][]{shimasaku01,nakamura03,weinmann09}.

Finally, Figure  \ref{MH2MHI}d shows that  \rmol\  is an increasing function ($r=0.25$) of \nuvr\ colour for galaxies with detections of both HI and CO. \citet{GASS1} reported a strong anti-correlation between HI mass fraction and colour. Figure \ref{fH2plots}d shows that there is also a fairly strong anti-correlation bewteen $f_{H2}$ and \nuvr\ colour. The fact that the anti-correlation of $f_{HI}$ with \nuvr\ appears to ``win'' probably reflects the fact that HI is dominant in regions such as outer galaxy disks where dust content is low and most of the starlight emitted by forming populations is emitted at UV wavelengths, whereas molecular gas tends to occur in the inner regions of galaxies where dust content is high and much of the light from young stars may be emitted in the infrared. In future work, we plan to look more carefully at these issues.

\section{Comparison with previous work}
\label{archive}

\begin{figure*}
\begin{minipage}{165mm}
\includegraphics[width=165mm]{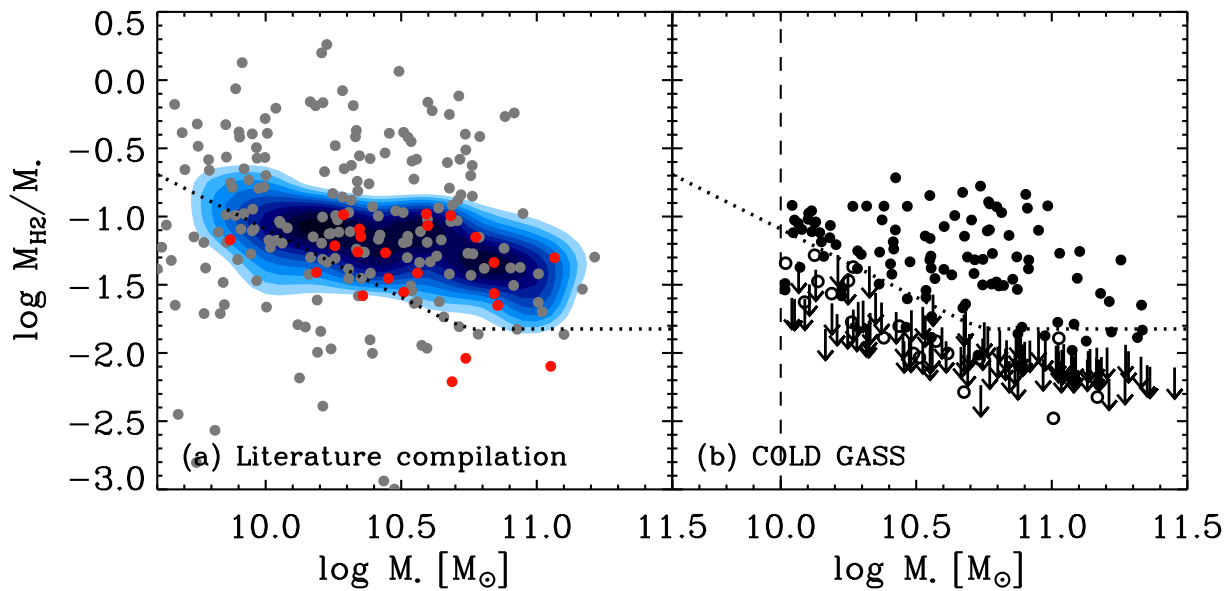}
\caption{{\it left:} Relation between \hmol\ mass fraction and stellar mass for a compilation of 263 galaxies with CO measurements available from the literature (gray filled circles).  The values of \mh\ have been homogenized for cosmology and the CO-to-\hmol\ conversion factor, but even with these corrections a scatter of more than two orders of magnitude is observed.  This is in strong contrast with predictions from the models of \citet{fu10} (contours), and smaller but more homogeneous datasets \citep[e.g.][shown as red circles]{leroy09}. {\it right:} In comparison, we show the same plot for the COLD GASS sample, with our minimum stellar mass selection (vertical dashed line) and our integration limit (dotted line) indicated. \label{figarch}}
\end{minipage}
\end{figure*}

To put the new COLD GASS results in context, we assembled CO data from the literature for 263 nearby galaxies in the SDSS survey. They are taken from the compilations of \citet{bettoni03}, \citet{yao03}, \citet{casasola04}, \citet{albrecht07}, \citet{komugi08} and \citet{obreschkow09}.  When multiple measurements are found for the same galaxy, the newest is assumed to supersede previous values. The values of \mh\ were then homogenized to the best of our ability using a common conversion factor ($X_{CO}=2.3\times10^{20}$ cm$^{-2}$ [K \kms]$^{-1}$) and cosmology ($H_0=70$ \kms\ Mpc$^{-1}$).  In addition, the SDSS photometry was reprocessed using the same technique used for the COLD GASS galaxies (see \S \ref{data}), and used to measure reliable and homogeneous stellar masses. 

In Fig. \ref{figarch}a, we show the relation between $M_{H2}/M_{\ast}$ and $M_{\ast}$ in this reference sample.  The molecular gas mass fraction has a spread of more than two orders of magnitude, and does not appear to correlate with stellar mass.  A significant fraction of this observed scatter can be attributed to measurement errors and inhomogeneities in the sample. The vast majority of these galaxies have $z<0.02$, and therefore tend to be significantly larger than the typical observing beam, resulting in important aperture problems.  Additional contributions to the artificially large scatter include different telescope calibrations, low $S/N$ detections and selection on IR luminosity, which tends to bias $f_{H2}$ high. 

As a comparison, we also show as contours in Figure \ref{figarch}a the relations produced by \citet{fu10} through semi-analytic modeling of galaxy formation including detailed prescriptions for the break-up of gas between the atomic and molecular phases.  The models predict a significantly smaller range in $M_{H2}/M_{\ast}$ than seen in the literature compilation.   But it is also clear that a systematically measured set of galaxies will produce more consistent results, for example the THINGS/HERACLES sample \citep{THINGS,leroy09}, which is also shown for comparison in Figure \ref{figarch}. 

The equivalent relation from COLD GASS is plotted in Figure \ref{figarch}b, showcasing the significantly reduced observational scatter compared to the literature compilation, but the increased dynamic range due to the rigorously measured upper limits for the CO non-detections. We note that the COLD GASS and HERACLES galaxies span a similar region in the plots.  The HERACLES galaxies provide resolved CO maps  for a much smaller sample of galaxies, so the two approaches are highly complementary. 

The main results presented in this paper are overall qualitatively similar to some earlier observations.  For example, \citet{sage93} found that $M_{H2}/M_{dyn}$ is independent of morphology,  just like we find $M_{H2}/M_{\ast}$ to be independent of concentration index.  However, they found that $M_{H2}/M_{HI}$ is a strong function of Hubble type \citep[see also][]{young89}, which we do not.   Based on these studies, we would have expected to find a significant population of early-type galaxies with large values of $M_{H2}/M_{HI}$.  We see no evidence for such systems in our sample.  

Much of the earlier work regarding CO in nearby galaxies focussed on the trends between $M_{H2}$, $M_{HI}$ and morphology, interaction state \citep[e.g.][]{braine93b}, or far infrared properties of the systems \citep[e.g.][]{sanders85}.  Using a set of galaxies still mostly based on the infrared-based FCRAO sample, \citet{bothwell09} however derive relations similar to ours, between gas fraction (atomic and molecular) and $B-$band luminosity.  Their findings are qualitatively similar to ours; they see that $M_{HI}/M_{\ast}$ decreases with luminosity, but $M_{H2}/M_{\ast}$ does not.  The breakthrough here is that COLD GASS allows us to {\it quantitatively} describe, in an unbiased sample, how the molecular gas component varies with several key physical parameters which are at the basis of the theoretical effort towards understanding the star formation process.  

\section{Summary}

We are conducting COLD GASS, a legacy survey for molecular gas in nearby galaxies. We target at  least 350 massive galaxies ($M_{\ast}>10^{10}$\msun) in the CO(1-0) emission line with the IRAM 30m telescope.  Because the survey is unbiased, it  will provide us with a complete view of the molecular gas properties of massive galaxies in the local universe, as well as the relations between molecular gas and other global galaxy properties.  The stellar mass and redshift ranges also ensure that we recover the total CO line flux of the galaxies with a single pointing of the IRAM 30m telescope, and that a single CO luminosity to $M_{H2}$ conversion factor is likely adequate. Finally, our observations  provide stringent upper limits on molecular gas fraction $M_{H2}/M_{\ast}<0.015$ in the case of CO non-detections. 

In this paper, we present  a catalog of CO(1-0) fluxes and $H_2$ masses (or upper-limits) for the first \ntot\ galaxies observed as part of COLD GASS, and report on their properties:

\begin{enumerate}

\item The detected molecular gas mass fractions ($M_{H2}/M_{\ast}$) 
are in the range of 0.9\% to 20\%, with a mean value 
$<M_{H2}/M_{\ast}>=0.066\pm0.039$. The mean gas mass fraction among the 
{\it detected} galaxies does not vary strongly 
with any global galaxy property except colour.

\item We detect the CO(1-0) line in $\sim50\%$ of COLD GASS 
galaxies, and while the detection rate is independent of 
stellar mass, it is a strongly decreasing function of stellar mass surface density,
concentration index and \nuvr\ colour. 
None of the \nred\ galaxies redder than \nuvr=5 were 
detected, and stacking them leads to a non-detection and a stringent 
upper limit of $<M_{H2}/M_{\ast}>=0.0016\pm0.0005$.

\item  The mean molecular gas mass fraction (averaged over galaxies
with detections and non-detections of the CO line)  is a roughly 
constant function of stellar mass, but a decreasing function 
of stellar mass surface density and concentration index. 
The observed trends are weaker than those observed by \citet{GASS1} 
for the atomic gas mass fraction in a similar sample of galaxies.  
Of all parameters investigated here, the molecular gas correlates 
most strongly with \nuvr\ colour, which is a tracer of specific star formation.

\item The molecular-to-atomic mass ratio, $R_{mol}$, has a mean value of $30\%$ 
over the entire sample.  It is a weakly increasing function of $M_{\ast}$ ($r=0.23$) and $\mu_{\ast}$ ($r=0.37$).

\end{enumerate}

One result that we wish to highlight is the existence of sharp thresholds in galaxy structural parameters such as stellar surface mass density, below which most galaxies have measurable atomic and molecular gas components, but above which the detection rate of both  the HI and CO lines drops drastically. This result was discussed previously for the HI in \citet{GASS1}. The fact that {\em the same sharp thresholds also apply to the CO} strongly suggests that onset of ``quenching'' processes in galaxies was associated with a change in their structure. We note that the same sharp drop in cold gas content is not seen as a function of stellar mass.  Intriguingly, atomic gas dominates in the minority of galaxies that are above threshold and that have significant cold gas content. One possible interpretation is that re-accretion of gas may still be possible following the quenching event.   In future work, we will examine galaxies on either side of the ``quenching threshold'' in more detail. We will also look more closely at the relationhip between molecular gas content and star formation.

\section*{Acknowledgments}

This work is based on observations carried out with the IRAM 30 m telescope. IRAM is supported by INSU/CNRS (France), MPG (Germany), and IGN (Spain).

We are grateful to Nario Kuno for providing us with total fluxes for his sample of nearby galaxies.   We wish to thank the staff of the IRAM observatory for their tremendous help in conducting our observations.  We thank the anonymous referee for a constructive and helpful report. 

RG and MPH are supported by NSF grant AST-0607007 and by a grant from the Brinson Foundation.



\appendix
\section{Spectral Gallery}
\label{atlas}

\subsection{Galaxies with CO line detections}

In Figures \ref{spectra_det}-\ref{lastspec_det}, we present SDSS imaging and IRAM spectra for the COLD GASS galaxies present in this data release for which we have securely detected the CO(1-0) line.  These are defined as those with $S/N>5$ in the CO line.  The tentative detections, those with $S/N<5$, are shown in Figures \ref{spectra_tent} and \ref{spectra_tent_last}.  All galaxies with a secure CO line detection have colours bluer that \nuvr=5.0, but cover uniformly the stellar mass range probed by COLD GASS. 

\begin{figure*}
\begin{minipage}{165mm} 
\includegraphics[width=165mm]{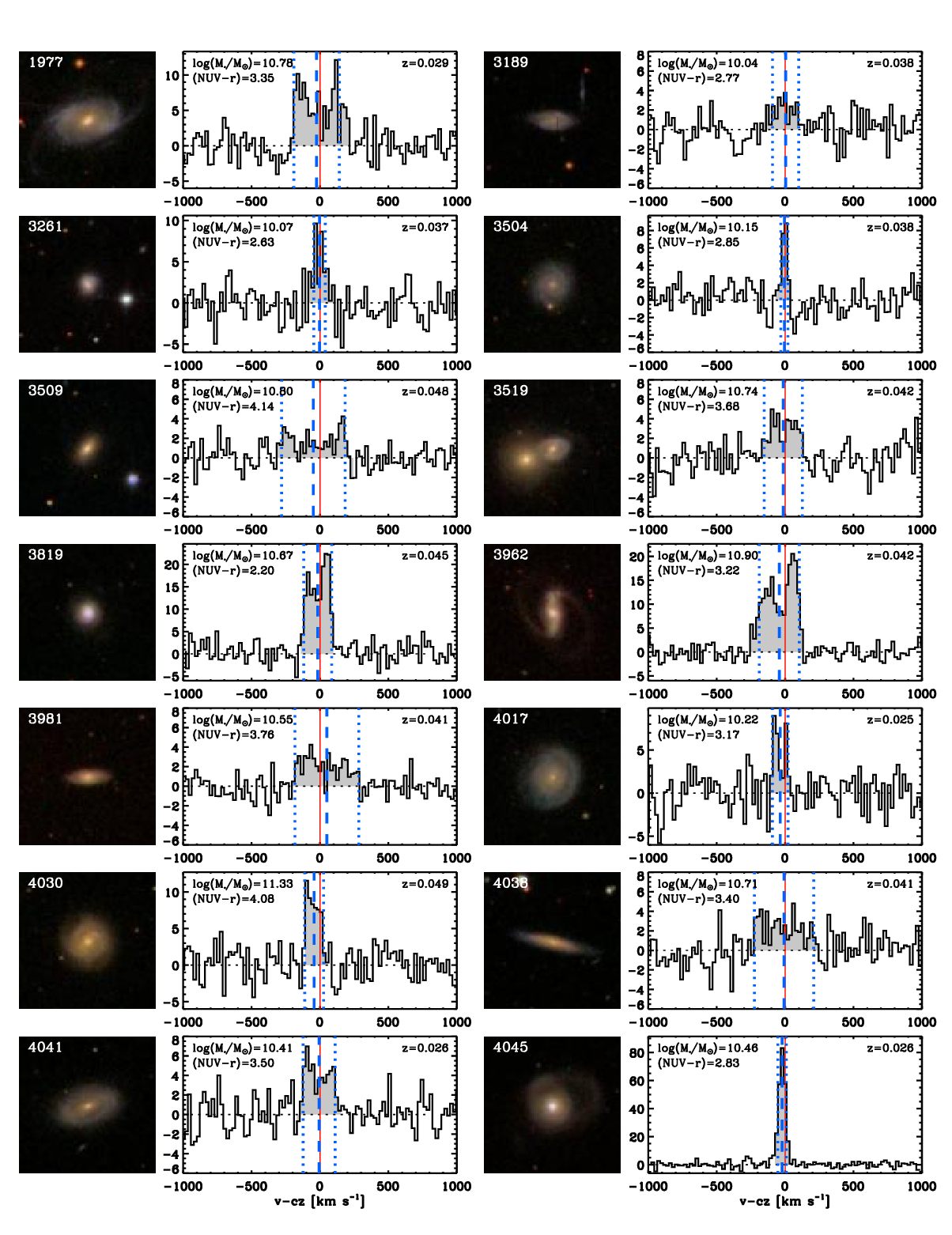}
\caption{SDSS three colour image ($1.5\arcmin \times 1.5\arcmin$) and CO(1-0) line spectrum of COLD GASS targets with a secure detection ($S/N>5$).   The solid red line shows the systemic redshift of the galaxy as determined from the SDSS fiber spectra.  The dashed blue line is the central velocity of the CO line, and the interval delimited by the two dotted blue lines is $W50_{CO}$, the full line width of the CO emission measured at half intensity. \label{spectra_det}}
\end{minipage}
\end{figure*}

\begin{figure*}
\begin{minipage}{165mm} 
\includegraphics[width=165mm]{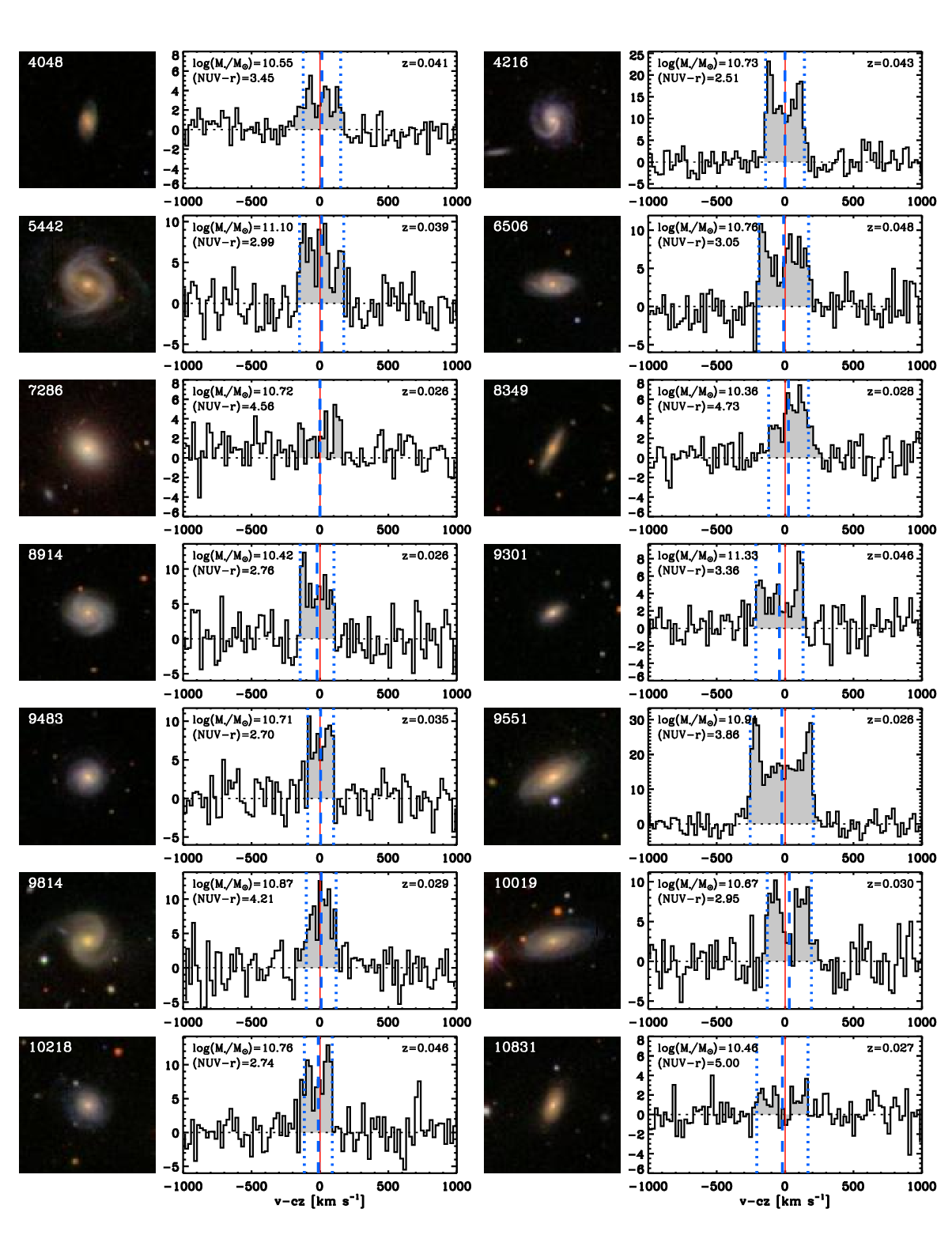}
\caption{continued from Figure \ref{spectra_det}}
\end{minipage}
\end{figure*}

\begin{figure*}
\begin{minipage}{165mm} 
\includegraphics[width=165mm]{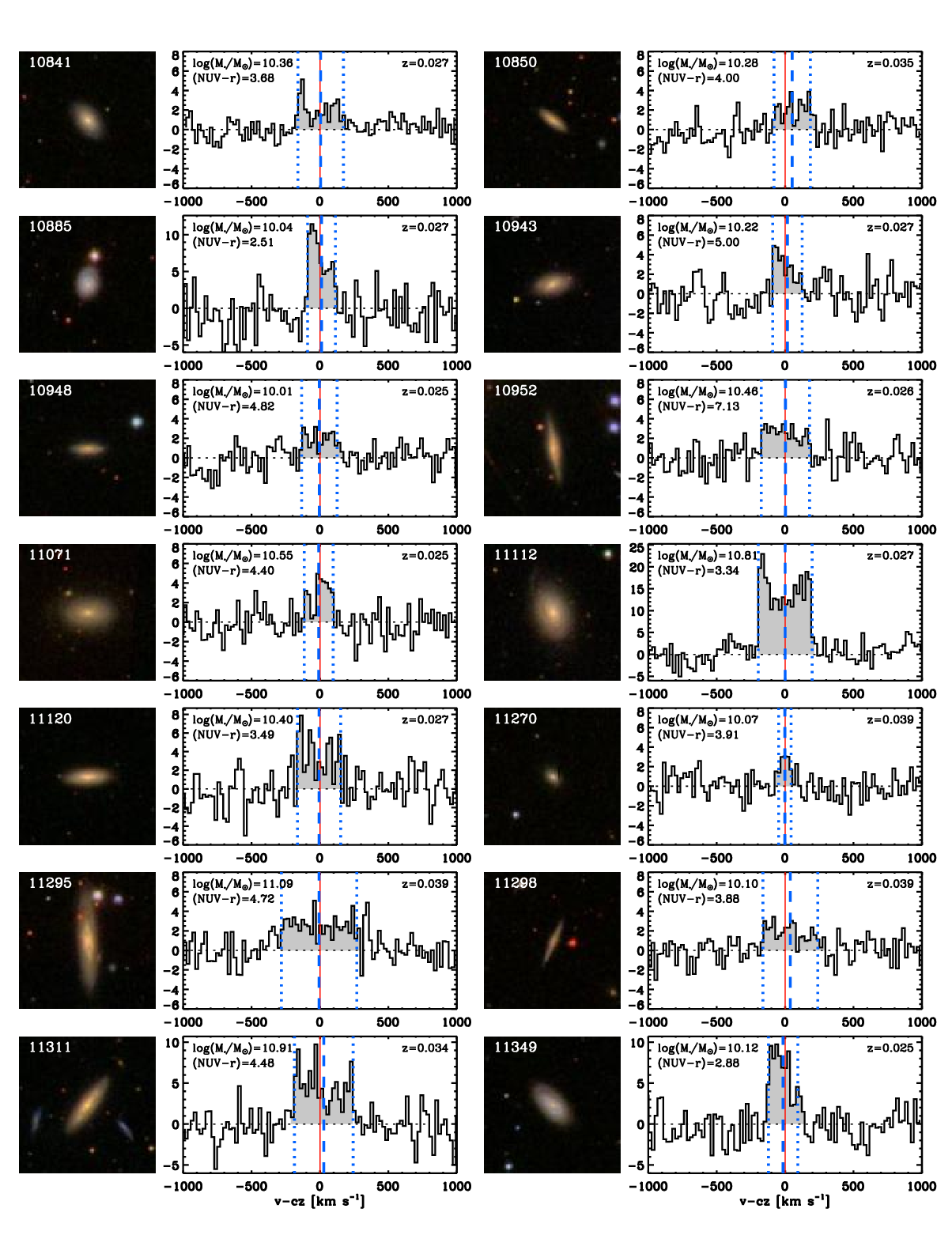}
\caption{continued from Figure \ref{spectra_det}}
\end{minipage}
\end{figure*}

\begin{figure*}
\begin{minipage}{165mm} 
\includegraphics[width=165mm]{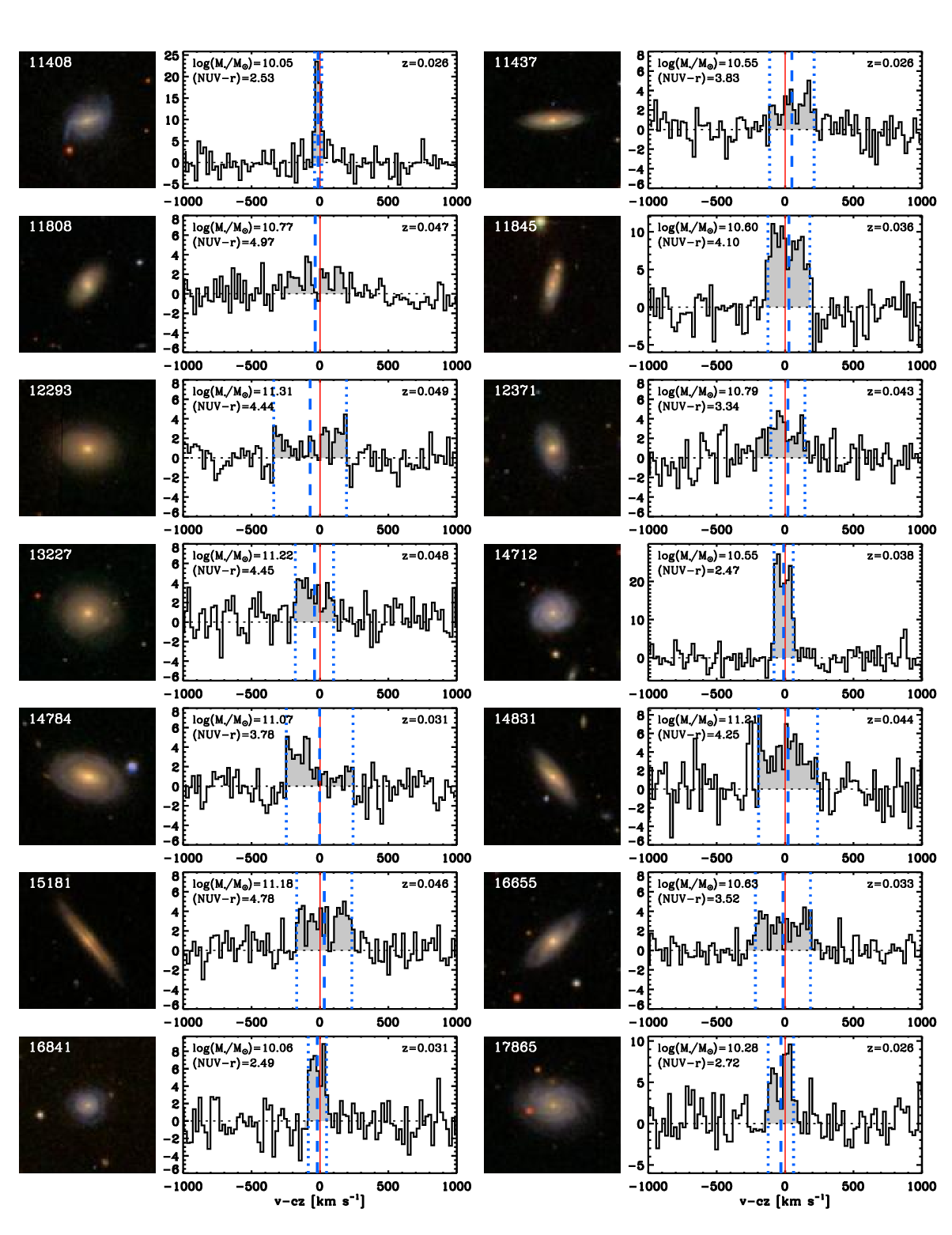}
\caption{continued from Figure \ref{spectra_det}}
\end{minipage}
\end{figure*}

\begin{figure*}
\begin{minipage}{165mm} 
\includegraphics[width=165mm]{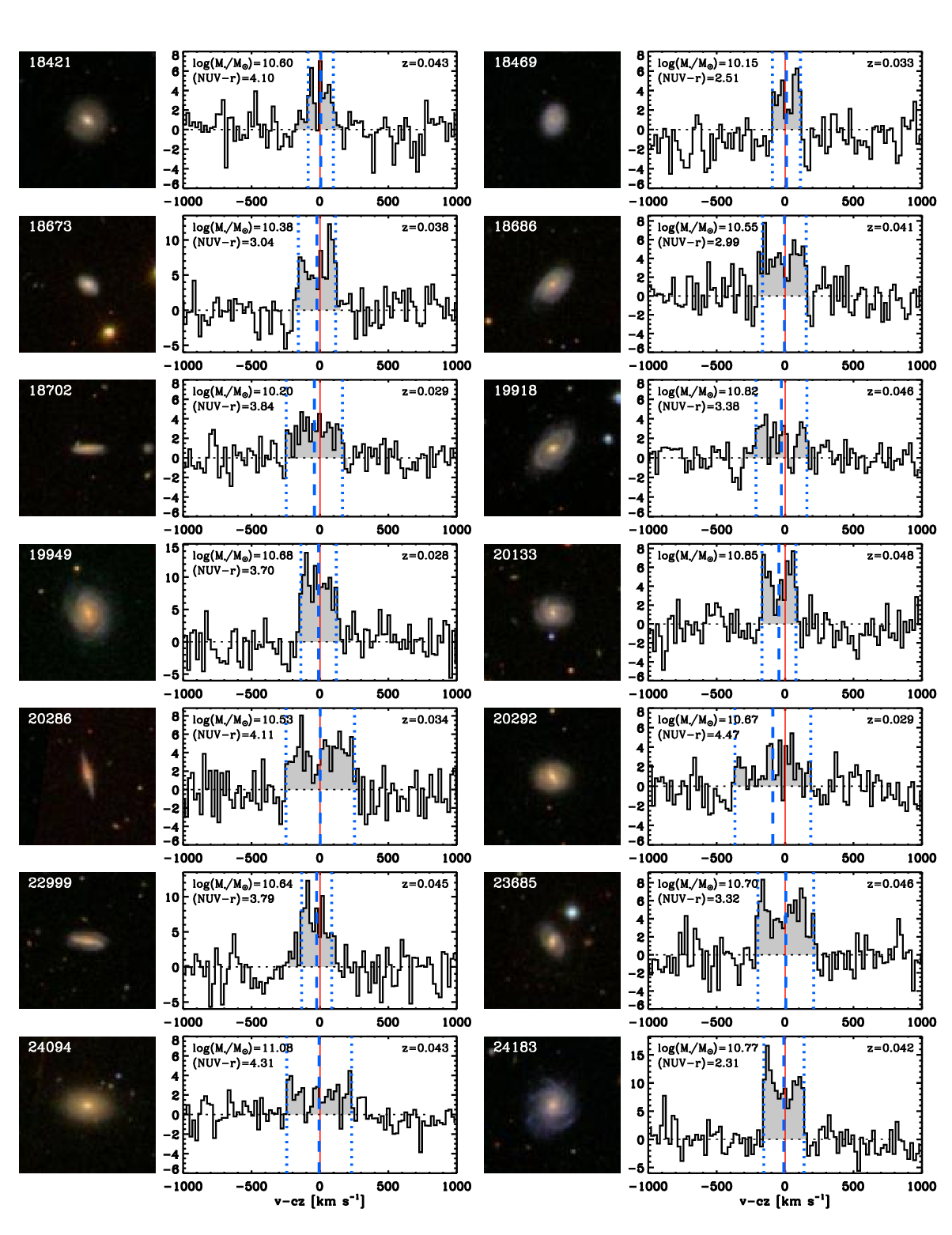}
\caption{continued from Figure \ref{spectra_det}}
\end{minipage}
\end{figure*}

\begin{figure*}
\begin{minipage}{165mm} 
\includegraphics[width=165mm]{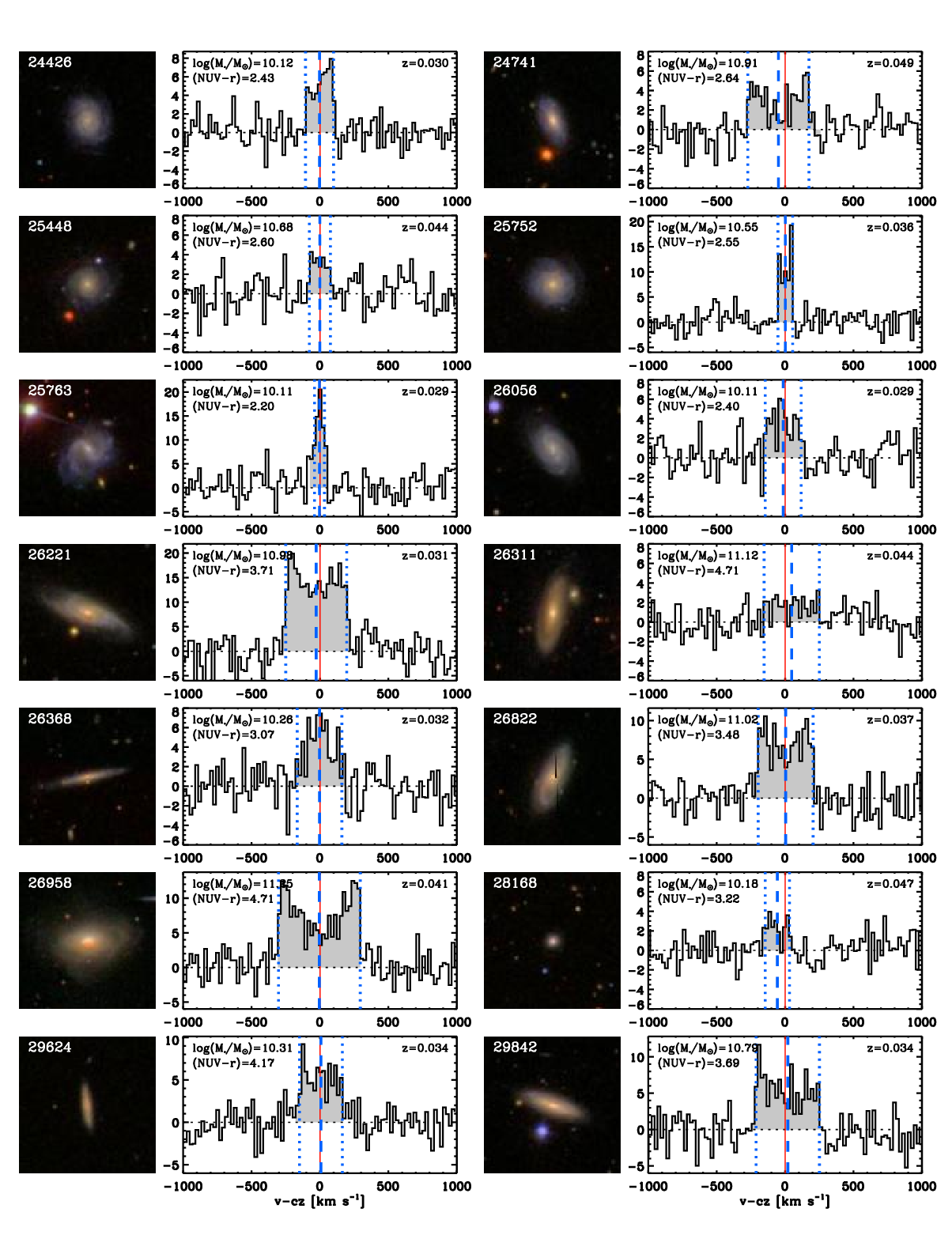}
\caption{continued from Figure \ref{spectra_det}}
\end{minipage}
\end{figure*}

\begin{figure*}
\begin{minipage}{165mm} 
\includegraphics[width=165mm]{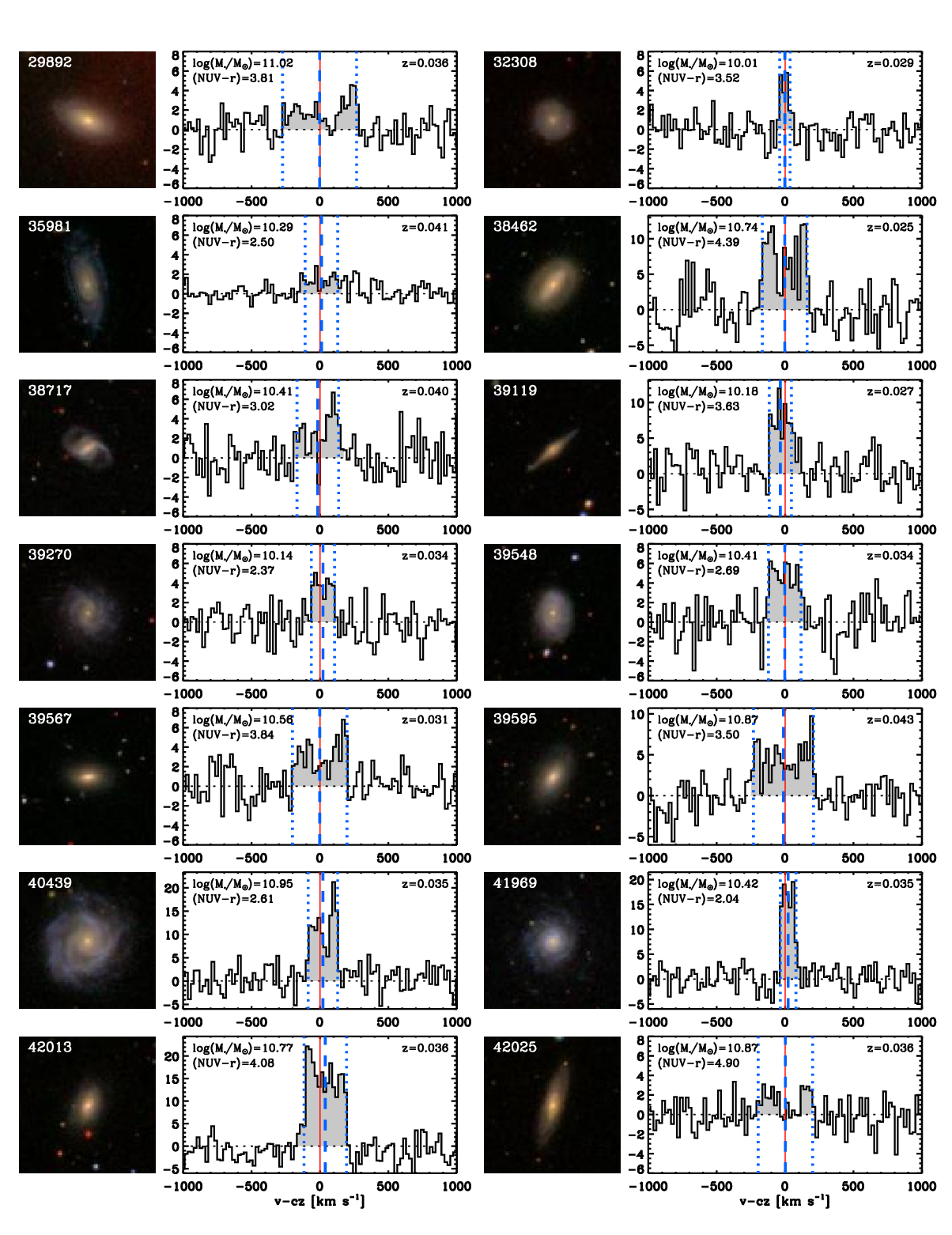}
\caption{continued from Figure \ref{spectra_det}}
\end{minipage}
\end{figure*}

\begin{figure*}
\begin{minipage}{165mm} 
\includegraphics[width=165mm]{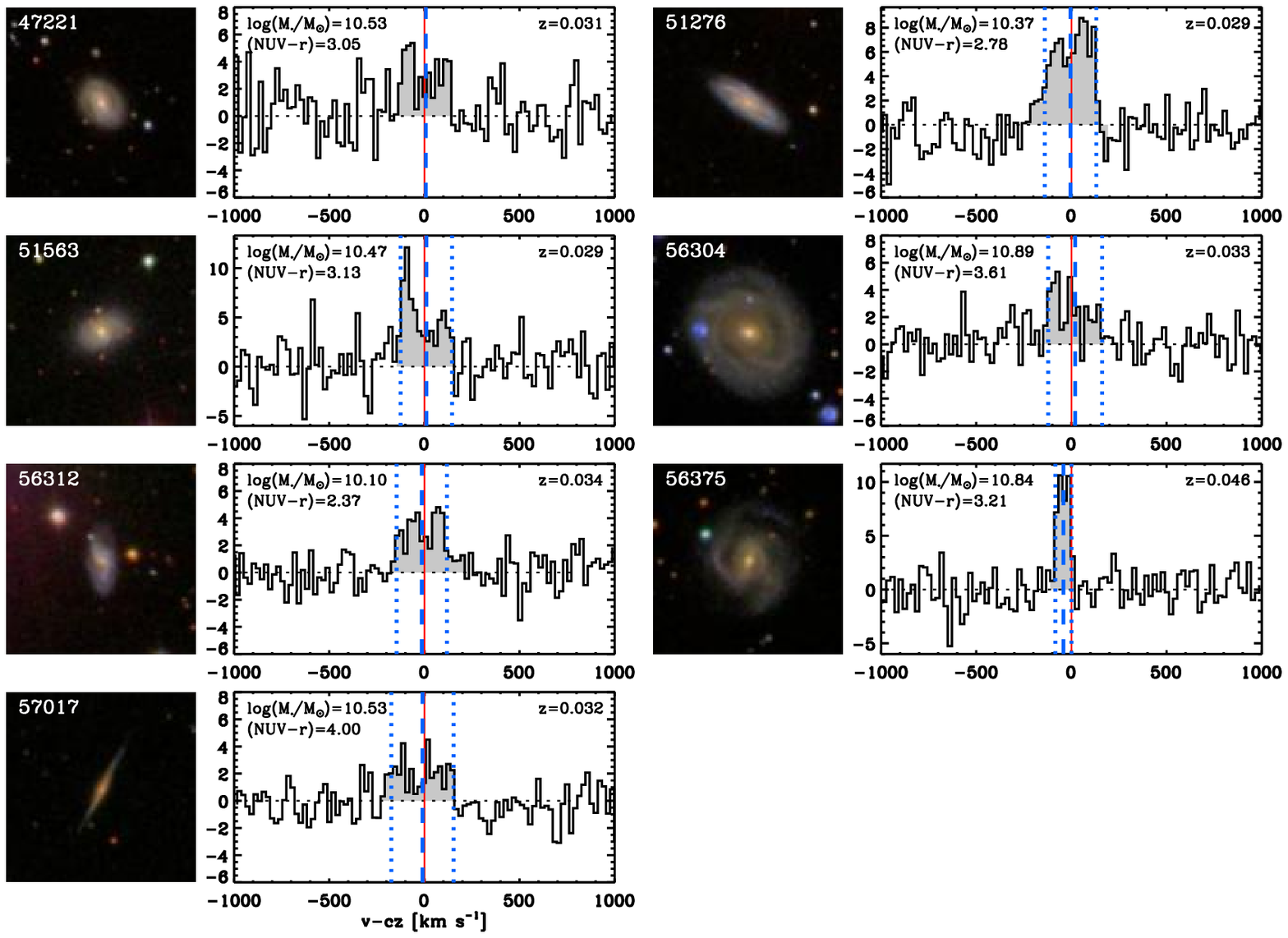}
\caption{end from Figure \ref{spectra_det}  \label{lastspec_det}}
\end{minipage}
\end{figure*}

\begin{figure*}
\begin{minipage}{165mm} 
\includegraphics[width=165mm]{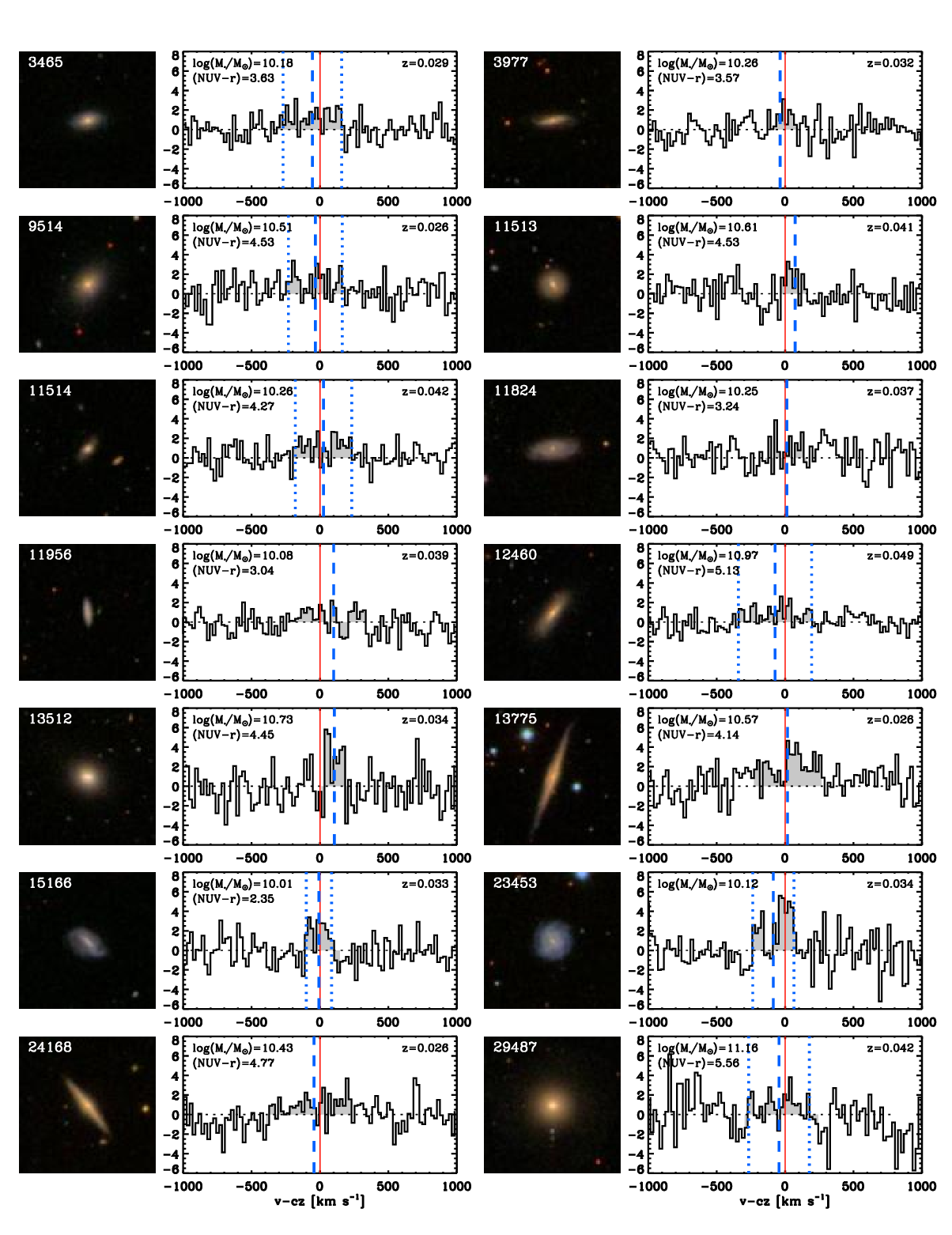}
\caption{SDSS three colour image and CO(1-0) line spectrum of COLD GASS targets with a tentative detection ($S/N<5$).  Lines are as described in the caption of Figure \ref{spectra_det}.  Given the low $S/N$ of these detections, the width determination is sometimes poor, and in these cases the corresponding vertical dotted lines are omitted. \label{spectra_tent}}
\end{minipage}
\end{figure*}

\begin{figure*}
\begin{minipage}{165mm} 
\includegraphics[width=165mm]{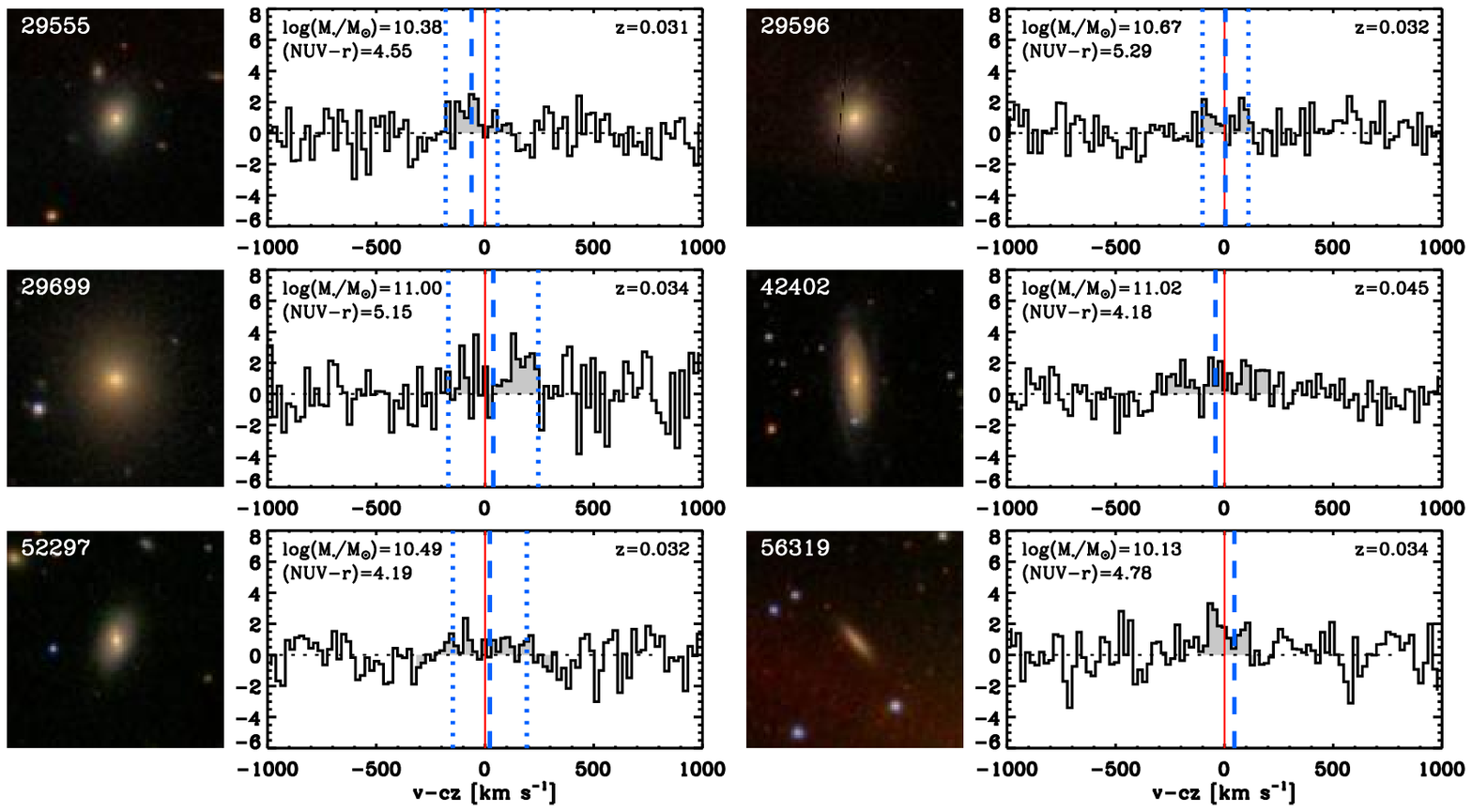}
\caption{end from Figure \ref{spectra_tent}.  \label{spectra_tent_last}}
\end{minipage}
\end{figure*}

\subsection{Galaxies with non-detection of the CO line}

For completeness, we show in Figures \ref{spectra_nondet}-\ref{spectra_nondet_last} the SDSS three-colour images of the CO non-detections.  These objects tend to be red (\nuvr$>5$), early-type looking ($C>2.6$) galaxies.

\begin{figure*}
\begin{minipage}{165mm} 
\includegraphics[width=165mm]{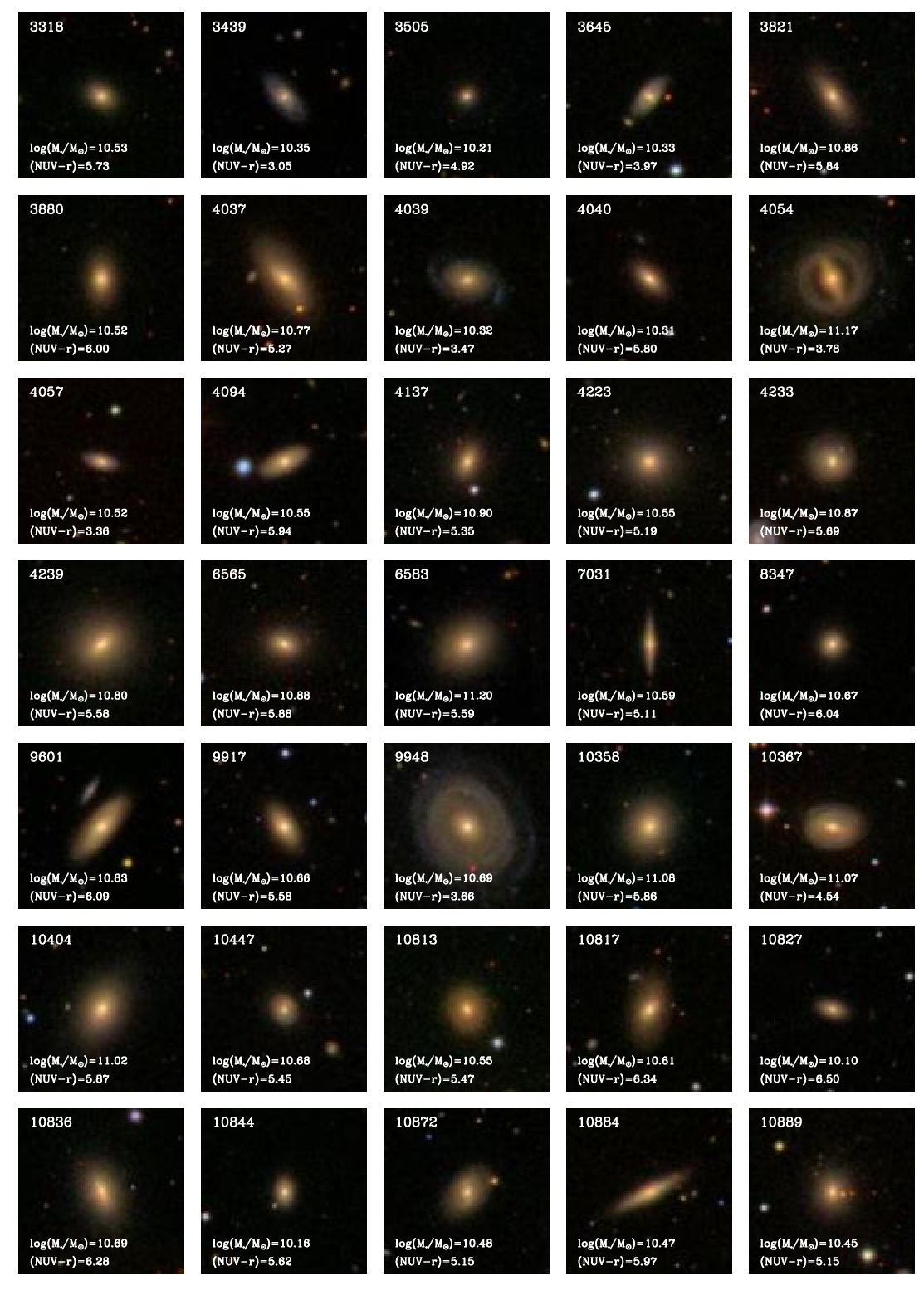}
\caption{SDSS three colour images of COLD GASS galaxies where the CO(1-0) line is not detected. Each image has a size of $1.5\arcmin \times 1.5\arcmin $ \label{spectra_nondet}}
\end{minipage}
\end{figure*}

\begin{figure*}
\begin{minipage}{165mm} 
\includegraphics[width=165mm]{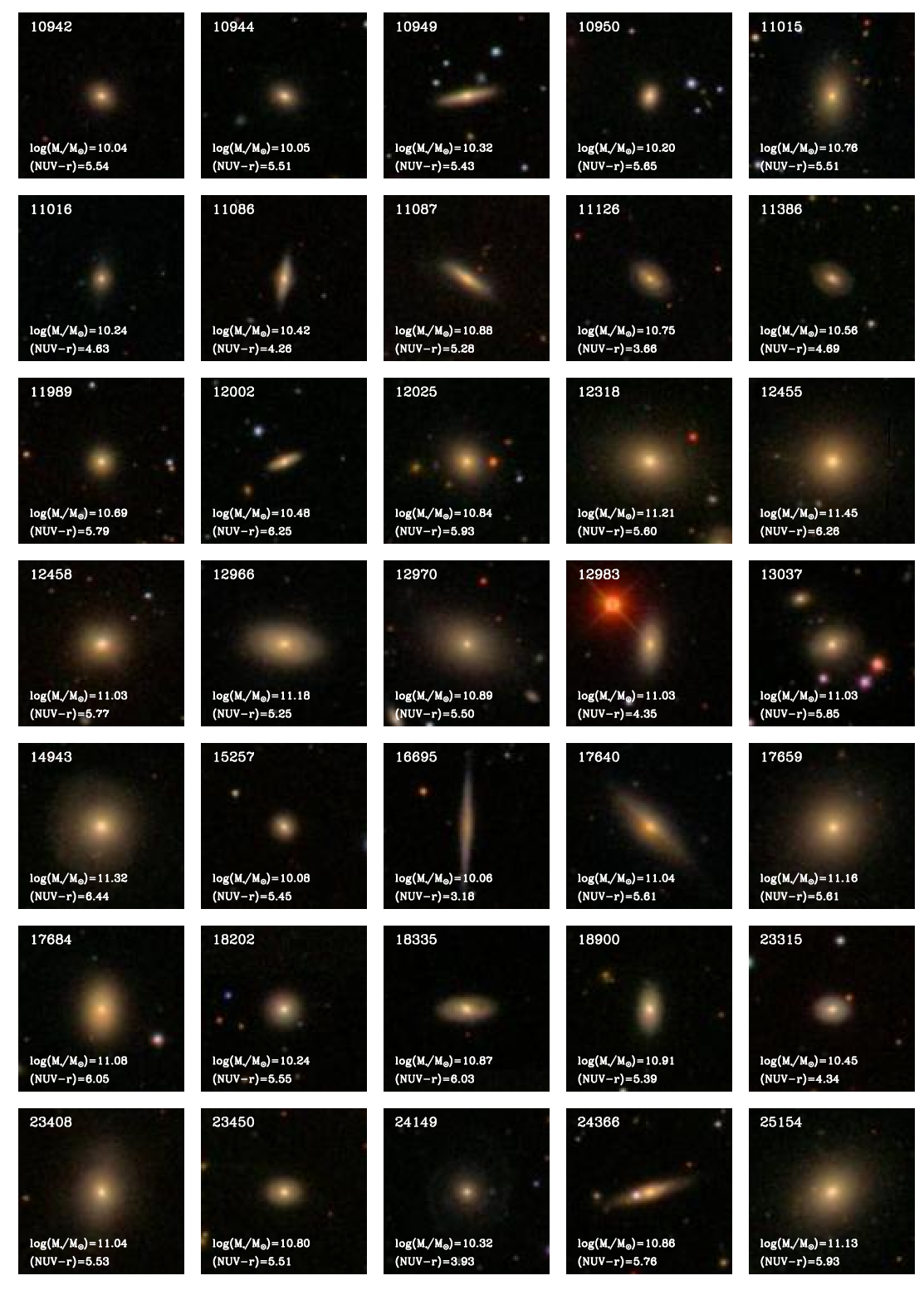}
\caption{continued from Figure \ref{spectra_nondet}}
\end{minipage}
\end{figure*}

\begin{figure*}
\begin{minipage}{165mm} 
\includegraphics[width=165mm]{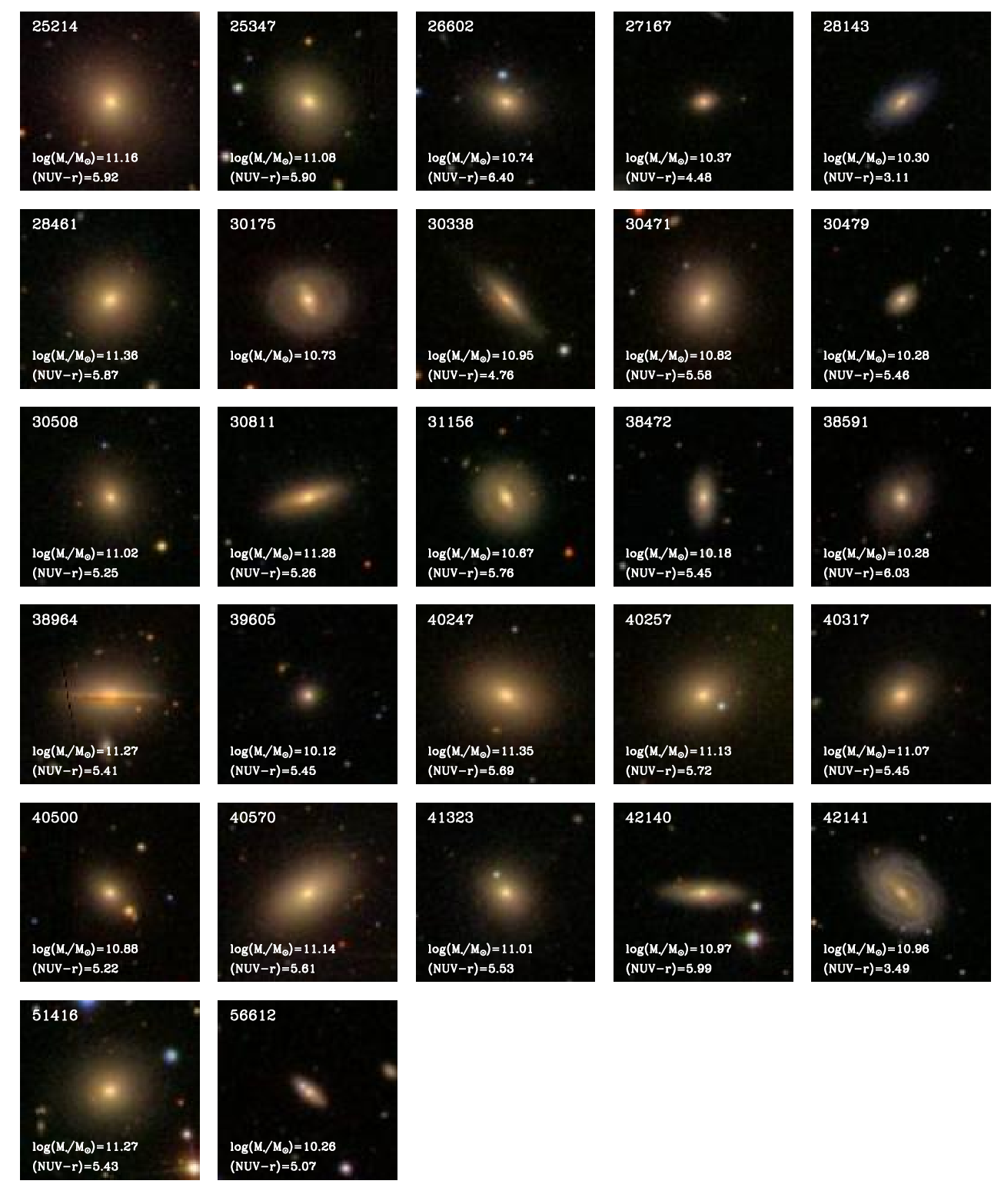}
\caption{end from Figure \ref{spectra_nondet} \label{spectra_nondet_last}}
\end{minipage}
\end{figure*}

\subsection{Offset pointings}

As described in \S \ref{offsets}, we perform additional off-center pointings for a small fraction of the COLD GASS sources. They are the larger galaxies ($D_{25}>40\arcsec$), with a strong detection in the central pointing.  These offsets are taken 16\arcsec (or three-quarters of the main beam) away from the centers.   In Figures \ref{spectra_off} and \ref{spectra_off2}, we show the SDSS images and IRAM spectra for the offset pointings.  The data reduction process is identical to that used for the central pointings and described in Section \ref{linemes}.  The emission lines are identified by examining the spectra at the expected positions based on the redshift on each galaxy, and the width of CO the line in the central pointing is used as additional information. Fluxes are measured by integrating over the region identified though this process, and the flux ratio between offset and central pointings is used to determine an appropriate aperture correction for each galaxy using the technique described in Section \ref{offsets}.

\begin{figure*}
\begin{minipage}{165mm} 
\includegraphics[width=165mm]{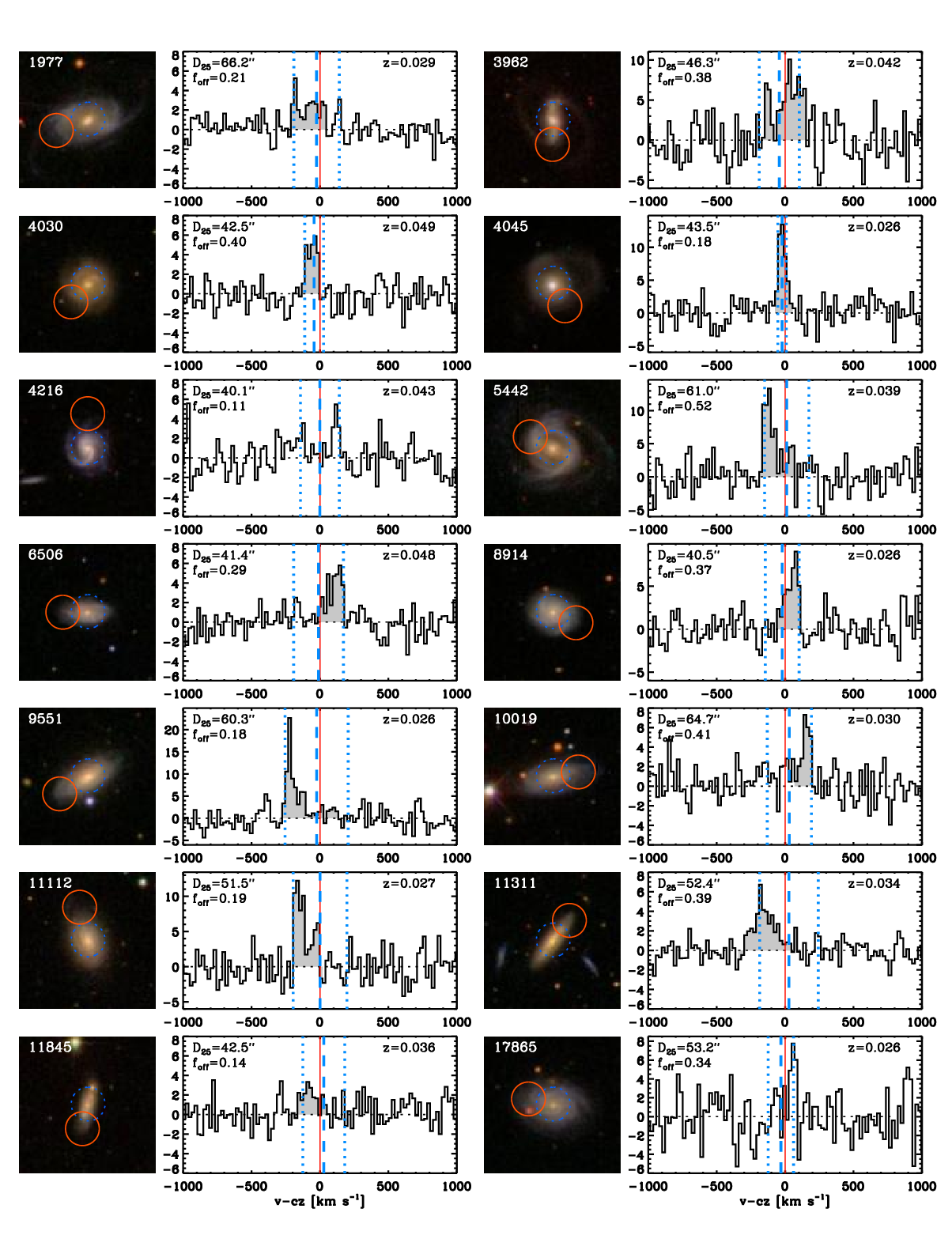}
\caption{SDSS three colour image and CO(1-0) line spectrum measured in offset pointings on COLD GASS galaxies.  The blue circle shows the central position of the 22\arcsec\ beam, and the orange circle the offset position, located three quarters of a beam from the center (for G1977, G4216, G9551, G11112 and G42013, the offset was taken one full beam away from the center, see \S \ref{offsets}).   On the spectra, we overplot the optical redshift of the galaxies (red lines), and the central position and widths of the line detected in the {\it central} pointings (blue lines).  The shaded region shows the region of the offset spectrum integrated to measure the line flux. The optical diameter from SDSS $g-$band imaging ($D_{25}$) and the flux ratio between offset and central pointings ($f_{off}$) are given in each case. 
\label{spectra_off}.}
\end{minipage}
\end{figure*}

\begin{figure*}
\begin{minipage}{165mm} 
\includegraphics[width=165mm]{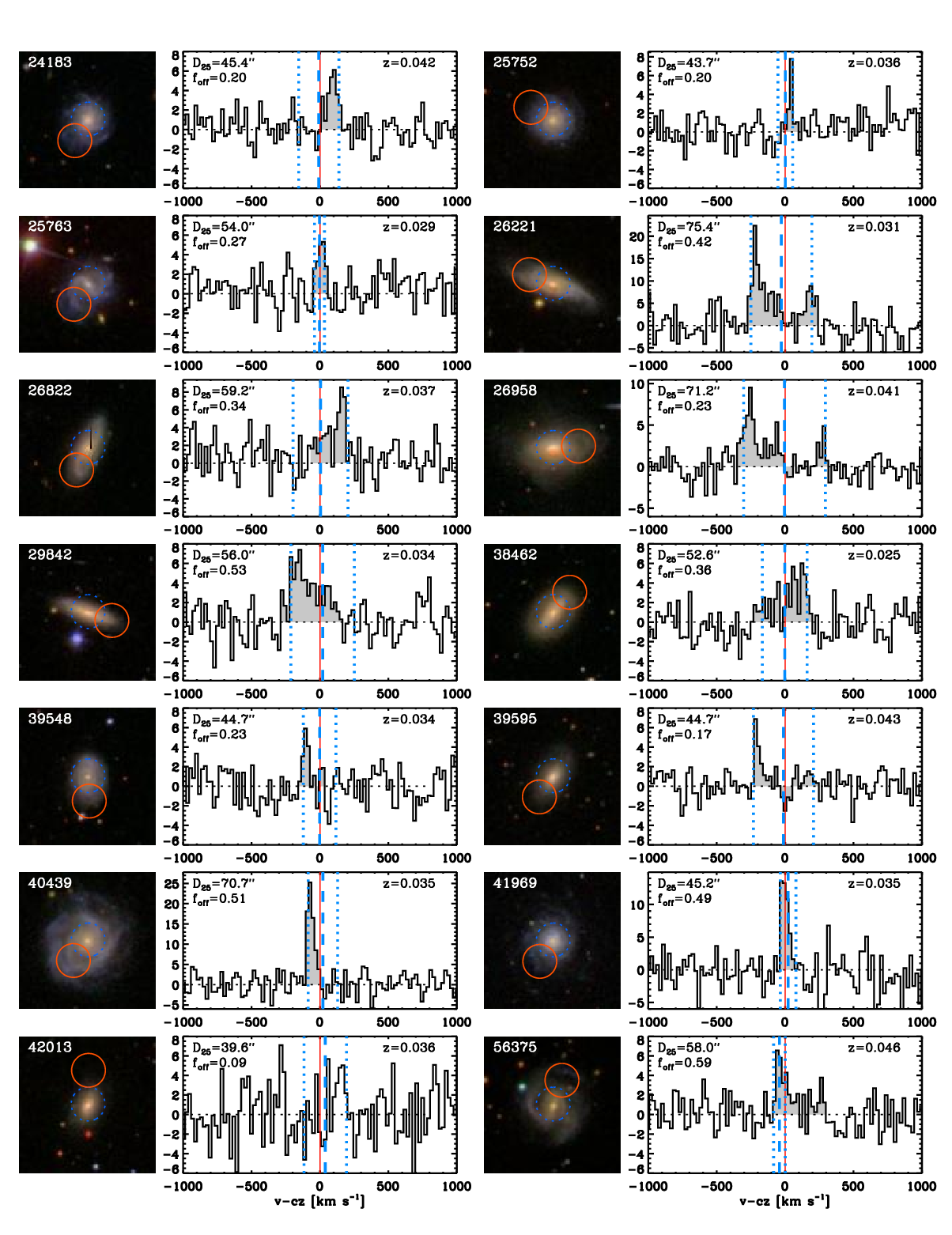}
\caption{continued from Figure \ref{spectra_off}. \label{spectra_off2}}
\end{minipage}
\end{figure*}

\end{document}